\pdfoutput=1

\documentclass[12pt]{article}
\usepackage
{amsmath,epsf,amssymb,latexsym,cite,shadow,eucal,theorem,enumerate,hhline}
\usepackage[matrix,arrow,frame,import,curve,color]{xy}

\usepackage{graphicx}
\newtheorem{theorem}{Theorem}

{\theorembodyfont{\rmfamily}}

\newif\iffigs\figstrue

\DeclareFontFamily{U}{rsf}{}
\DeclareFontShape{U}{rsf}{m}{n}{
  <5> <6> rsfs5 <7> <8> <9> rsfs7 <10-> rsfs10}{}
\DeclareMathAlphabet\Scr{U}{rsf}{m}{n}

\def\O{\Scr{O}}
\def\C{{\mathbb C}}
\def\P{{\mathbb P}}

\def\Z{{\mathbb Z}}

\def\sHom{\operatorname{\Scr{H}\!\!\textit{om}}}

\def\Ext{\operatorname{Ext}}

\def\End{\operatorname{End}}

\def\coker{\operatorname{coker}}
\def\supp{\operatorname{supp}}
\def\neg{\operatorname{neg}}
\def\Spec{\operatorname{Spec}}

\def\SO{\operatorname{SO}}

\def\SU{\operatorname{SU}}
\def\GU{\operatorname{U{}}}

\def\id{{\mathbf{1}}}

\def\Wt{\widetilde W}
\def\lqs#1#2#3{{\left( {\begin{smallmatrix}#1\\#2\;\; #3\end{smallmatrix}}\right)}}
\def\p{\partial}

\def\CY{Calabi--Yau}
\def\LG{Landau--Ginzburg}

\def\cA{{\Scr A}}

\def\cE{{\Scr E}}
\def\cF{{\Scr F}}

\def\DC{\mathbf{D}}

\def\ff#1#2{{\textstyle\frac{#1}{#2}}}
\def\half{\frac{1}{2}}

\def\poso#1{#1\save="x"!LD+<0pt,-0.5mm>;
  "x"!RD+<0pt,-0.5mm>**\dir{.}\restore}

\def\pz{\phantom{-}}

\def\eqn#1#2{\begin{equation}#2
  \ifx{#1}{}\else\label{#1}\fi\end{equation}}

\def\smc#1{{\scriptstyle\C^{#1}}}

\def\GUL{\GU(1)_{\text{L}}}

\def\rep#1{{{\boldsymbol{#1}}}}
\def\brep#1{{{\overline{\boldsymbol{#1}}}}}

\newcommand\Qb{\overline{Q}}

\newcommand\gammab{\overline{\gamma}}
\newcommand\cb{\overline{c}}
\newcommand\qb{\overline{q}}
\newcommand\Xb{\overline{X}}
\newcommand\Jb{\overline{J}}
\newcommand\xb{\overline{x}}
\newcommand\psib{\overline{\psi}}
\newcommand\nut{\widetilde{\nu}}
\newcommand\htld{\widetilde{h}}
\def\GE{\operatorname{E{}}}
\def\ra{\rangle}
\def\la{\langle}
\def\ep{\epsilon}

\begin{document}

\begin{titlepage}
\begin{flushright}
AEI-2010-120\\
August 2010
\end{flushright}
\vspace{.5cm}
\begin{center}
\baselineskip=16pt
{\fontfamily{ptm}\selectfont\bfseries\huge
(0,2) Elephants\\[20mm]}
{\bf\large  Paul S.~Aspinwall$^1$, Ilarion V.~Melnikov$^2$ 
and M.~Ronen Plesser$^1$
 } \\[7mm]

{\small

$^1$Center for Geometry and Theoretical Physics, 
  Box 90318 \\ Duke University, 
 Durham, NC 27708-0318 \\ \vspace{10mm}

$^2$Max-Planck-Institut f\"ur Gravitationsphysik (Albert-Einstein-Institut),\\
Am M\"uhlenberg 1, D-14476 Golm, Germany\\ \vspace{6pt}

 }

\end{center}

\begin{center}
{\bf Abstract}
\end{center}
We enumerate massless $\GE_6$ singlets for (0,2)-compactifications of
the heterotic string on a \CY\ threefold with the ``standard
embedding'' in three distinct ways. In the large radius limit of the
threefold, these singlets count deformations of the \CY\ together with
its tangent bundle.  In the ``small-radius'' limit we apply \LG\
methods.  In the orbifold limit we use a combination of geometry and
free field methods. In general these counts differ. We show how to
identify states between these phases and how certain states vanish
from the massless spectrum as one deforms the complex structure or
K\"ahler form away from the Gepner point.  The appearance of extra
singlets for particular values of complex structure is explored in all
three pictures, and our results suggest that this does not depend on 
the K\"ahler moduli.

\end{titlepage}

\vfil\break


\section{Introduction}    \label{s:intro}

The earliest form of model building in string theory consisted of
``embedding the spin connection in the gauge group'' for a \CY\
compactification of the $\GE_8\times\GE_8$ heterotic string
\cite{CHSW:}. These correspond to (0,2)-compactifications that happen
to have $N=(2,2)$ worldsheet supersymmetry. A natural question to ask
concerns the counting of massless states in uncompactified spacetime
which are singlets under the unbroken $\GE_8\times\GE_6$ gauge symmetry. This
turns out to be a fascinating question that has received rather sporadic
attention in the past 25 years.

These massless states correspond to first order deformations of the theory.
Marginal deformations must preserve the (0,2) superconformal
symmetry~\cite{Banks:1988yz}.  Deformations that preserve
the full (2,2) invariance constitute the familiar (2,2) moduli space.  Its dimension
is constant~\cite{Dixon:}, and in a geometric phase it corresponds to the
unobstructed~\cite{Tian:def} deformations of complex structures and
changes in the complexified K\"ahler form.  The remaining moduli that
only preserve (0,2) invariance are harder to describe.  As a first step, we
may count the massless gauge singlets in the four-dimensional
effective theory.  Each of these is a first order deformation that may be
obstructed at higher order.

Unfortunately, the identification of all massless singlets at a generic point
in the (2,2) moduli space is well beyond our current abilities.  To make 
progress, we must work at certain limiting points where the spectrum is
accessible to available techniques.  These include large radius points,
\LG\ loci and the Gepner points they contain, and orbifolds.  In each of
these points the techniques used to identify the singlets are rather different,
and the resulting description of the space of first order deformations might
appear as mysterious as an elephant to the group of proverbial blind 
men from Indostan.  Can these different descriptions be reconciled?

In the large radius limit of a \CY\ phase, the counting of the singlets is 
quite easy to
visualize. The (2,2) singlets manifest themselves as infinitesimal
deformations of the complex structure or complexified K\"ahler form
of the \CY, while the less familiar (0,2) singlets correspond to
first order deformations of the tangent bundle, counted by $H^1(\End T).$ 
This group can jump with complex structure \cite{Berglund:1990rk}; 
moreover a ``generic''
first order (0,2) deformation is expected to be lifted by world-sheet
instantons~\cite{DSWW:}.  One might expect, therefore,
that the Gepner models corresponding to certain \CY\
threefolds might count the number of (0,2)-deformations
differently. After all, the Gepner model describes physics
at some ``minimal radius'' for the \CY\ threefold, well away from the large
radius limit, for a special choice of complex structure.
Singlet counts for Gepner models were comprehensively listed 
in~\cite{Lutken:1988hc}.

So perhaps the different aspects of the elephant cannot be
reconciled. It may be that the number of singlets varies wildly across
the moduli space. We will argue here that is not the
case. The behaviour of the singlets is quite orderly, with relatively
modest jumping in the singlet count.   As is well
studied, the quintic threefold provides a remarkably boring case study,
where the number of singlets is fixed except for a handful of singlets
associated with extra $\GU(1)$ gauge symmetries at the Gepner point. 

In the case of the quintic, the gauged linear sigma model offers a 
beautiful explanation of this behaviour.  As already noted in~\cite{W:phase},
the (2,2) GLSM describing a \CY\ complete intersection in a toric variety
has natural (0,2)-preserving deformations encoded in a (0,2) superpotential
for the gauge theory.  The holomorphic parameters of this superpotential
encode the ``toric'' K\"ahler moduli, the ``polynomial'' complex structure
moduli, as well as a subset of classically unobstructed bundle moduli.
This GLSM parameter space was recently studied in some detail in the
case of \CY\ hypersurfaces~\cite{Kreuzer:2010ph}.  Remarkably, 
these GLSM
deformations have been argued to correspond to exactly marginal
deformations of the (0,2) 
theory~\cite{Silverstein:1995re,Basu:2003bq,Beasley:2003fx}.

Getting back to the quintic, it is not hard to see that all elements of
$H^1(\End T)$ can be represented as deformations of the (0,2) GLSM
superpotential.  Hence, it is not too surprising that the singlet spectrum
at the Gepner point simply differs by a few states associated to the 
un-Higgsing of additional gauge symmetries.  

More generally,  there are certainly
models where $H^1(\End T)$ is not fully described by the (0,2) GLSM, and
the additional singlets, unprotected by GLSM arguments, should suffer the
fate of the ``generic'' large radius singlet and become massive away from
the large radius limit.

In this paper we will study various cases which have a little more
structure than the quintic. We will show how the bulk of the spectrum
stays fixed and can be tracked nicely between the \CY\  and
\LG\ pictures. We will also see how various massless states can 
appear in some subspaces of the moduli space. In some cases 
these extra states can be tracked all the way from the Gepner
model to the large radius limit.

The orbifold is an important intermediate step on a path
from the Gepner model to the large radius limit. Comparing the
orbifold to the large radius limit is extremely well studied in the
context of (2,2) theories. Here the relationship between the orbifold
and its resolution is now generally known as the McKay
correspondence. A McKay correspondence for the (0,2)-case has been
quite neglected, despite its origin in string theory being as old
\cite{DHVW:}. We make some first efforts in this direction here.  In
particular, at the orbifold limit the massless states can be
characterised as ``untwisted'' or ``twisted,'' and we are able to
compute the spectrum of states of both types.  In the case of a \CY\
space with a curve of quotient singularities this involves
understanding how the six-dimensional theory determined by the
quotient is compactified on the singular curve and leads to a twisted
compactification familiar from the study of wrapped D-branes.
This is a first step toward a McKay correspondence, but there are
some subtleties we do not resolve here.

We will focus on 4 examples of \CY\ threefolds in a (weighted)
projective space, each of which has its own merit:
\begin{itemize}
\item a quintic in $\P^4$ is the simplest and most studied case;
\item a sextic in $\P^4_{21111}$ has extra singlets states at small radius;
\item a septic in $\P^4_{31111}$ is a blown-up orbifold and
demonstrates the (0,2) McKay correspondence;
\item an octic in $\P^4_{22211}$ exhibits many complications,
including extra singlets appearing both at special radii and at special
complex structure.
\end{itemize}

The \LG\ locus for the sextic and octic theories has additional singlets
in comparison to the large radius computation, which are not associated
with an enhanced gauge symmetry.  What is the fate of these singlets
as we move away from the \LG\ locus by turning on a K\"ahler deformation?
The only reasonable possibility is that they acquire a K\"ahler-dependent
mass term, which is indeed allowed by the quantum symmetry of the
\LG\ orbifold and consistent with the fact that the number of
additional chiral singlets is even.

This would be challenging to verify directly even at the Gepner point, 
since it would require us to 
compute correlators of several twisted states.  Luckily, we have a tool
at our disposal that would be singularly unhelpful to the six blind men:
we can take a look at our elephant in the mirror.  Using mirror symmetry
we are able to show that the extra singlets do indeed acquire a 
K\"ahler-dependent mass.

The extra singlets provide examples of states with K\"ahler
dependent masses; however we observe that in all of our examples
every large radius singlet, whether it is a (0,2) 
GLSM deformation or not, remains massless at the \LG\ locus.  Thus, we
have yet to find an example of a ``generic'' large radius singlet that is lifted 
by world-sheet instantons.  Instantons
could well lead to higher order obstructions for these first order (0,2)
deformations, but we find it remarkable that an
instanton-induced mass term for the non-GLSM singlets, while allowed
by symmetries, is not generated.
This suggests that there may be a non-renormalization theorem with a 
wider applicability than the one currently known for the subspace of GLSM
deformations.\footnote{We have a hint of this possibility already from the
(2,2) moduli space:  a generic \CY\ hypersurface will have non-toric and
non-polynomial deformations that are not described by familiar holomorphic
couplings in the GLSM Lagrangian;  although standard GLSM arguments
do not apply to these, each such deformation corresponds to a deformation
of the CFT.}

The rest of the article is organized as follows:  in sections \ref{s:LG} and 
\ref{s:T} we review and develop the technology necessary to study
heterotic spectra in \LG\ and \CY\ phases.  Section \ref{s:compare} is
devoted to a comparison of the general results, while section \ref{s:eg}
contains specific computations in the examples.


\section{Singlet Spectrum at the Landau--Ginzburg Locus} \label{s:LG}
Describing the massless spectrum of a heterotic vacuum as a function
of the moduli is a difficult affair even in string perturbation
theory, since it requires a knowledge of the marginal operators in
a non-trivial SCFT.  At certain points in the moduli space the SCFT
may reduce to a solvable theory: for instance, it might be an orbifold
of a free theory or a Gepner model.  When this holds, the full
perturbative string theory is under control: the spectrum is
computable, and any scattering amplitude can be reduced to
an integral over the moduli space of a punctured Riemann surface.  In
principle, conformal perturbation theory can then be used to determine
these properties in an open neighborhood of the solvable point.
Unfortunately, in practice conformal perturbation theory is difficult
to carry out in full generality.  For instance, even in an orbifold of
$T^6$ little is known about the dependence of spectrum and amplitudes
on the twisted sector moduli corresponding to K\"ahler deformations.

To make progress it turns out to be useful to sacrifice a little of
the ambition: if we cannot determine the spectrum as a function of all
the moduli, perhaps we can do so on some suitably nice locus in the
moduli space.  For instance, in the example of a $T^6$ orbifold we can
look at the dependence of the spectrum on all of the untwisted moduli.
This example is perhaps not very exciting, since as far as this
dependence is concerned, we are basically dealing with a solvable
theory.  The \LG\ models provide an important class of examples where
the massless spectrum can be determined, even though the theory is not
a solvable SCFT.  In particular, the \LG\ description allows us to
follow the massless spectrum as parameters in the superpotential are
varied.

\subsection{(2,2) \LG\  Generalities} \label{ss:lggen}
In this section we review some standard results on (2,2) Landau-Ginzburg models~
\cite{Mart:,VW:} and their uses in heterotic compactification~\cite{Kachru:1993pg}.
A (2,2) Landau-Ginzburg theory with a UV R-
symmetry is defined by a Lagrangian for $N$ chiral superfields $X^i$ with canonical 
kinetic terms and a quasi-homogeneous superpotential $W(X)$ satisfying
\begin{equation*}
W(\lambda^{\alpha_i} X^i) = \lambda W(X).
\end{equation*}
The superpotential coupling is a relevant deformation of the free
theory, and under suitable conditions the IR fixed point is believed
to be a non-trivial compact (2,2) SCFT, with the UV R-symmetry
corresponding to the R-symmetry of the IR theory.  In such theories
the critical point set of $W$, i.e. points where $dW = 0$, is
the origin in $\C^n$, and without loss of generality the $\alpha_i$
may be taken to be $0 < \alpha_i \le \ff{1}{2}$.

While following the \LG\ RG flow is a non-trivial affair, it turns out
that a number of properties of the theory, in particular those
involving the supersymmetric ground states, are independent of the RG
details.  Perhaps the most elegant way to encapsulate the accessible IR
physics is via a representation of the left-moving $N=2$
super-conformal algebra in the cohomology of the right-moving
supercharge $\Qb$ of the UV theory~\cite{Witten:1993jg}.  The
propagating fields in a (2,2) chiral multiplet $X^i$ and its complex
conjugate anti-chiral multiplet $\Xb^i$ consist of the bosonic $x^i$,
its complex conjugate $\xb^i$, the right-moving fermions $\psi_+^i,
\psib_+^i$, and the left-moving fermions $\gamma_-^i,\gammab_-^i$.
Define $T, J, G^\pm$ by\footnote{We
  use the notation where $\pm$ denotes R-charge, not world-sheet
  chirality.}
\begin{align}
\label{eq:Tj}
T^0 & =  \sum_i \left\{-\gamma^i \p \gammab^i -2\p\xb^i\p x^i \right\},\nonumber\\
J     & =  \sum_i\left\{ (\alpha_i-1) \gamma^i \gammab^i -2 \alpha_i x^i \p\xb^i\right\},
\nonumber\\
T     & =  T^0 - \ff{1}{2} \p J, \nonumber\\
G^- & = 2i\sqrt{2} \sum_i \gamma^i\p\xb^i, \nonumber\\
G^+    & = i\sqrt{2} \sum_i \left\{(1-\alpha_i)\gammab^i\p x^i -\alpha_i x^i \p
\gammab^i \right\}.
\end{align}
It is easy to show that these operators are $\Qb$-closed up to the equations of
motion of the UV theory; moreover, the super-renormalizability of the Lagrangian
allows us to evaluate the algebra of these operators via free-field OPEs.  The result
is that this is a representation of the $N=2$ algebra with central charge
$c = 3\sum_i(1-2\alpha_i)$.  The conformal weights $h$  and charges $q$ of the
fundamental fields are given in table~\ref{table:charges}; the table also lists the
charges $\qb$ under the right-moving R-symmetry.

\begin{table}[t]
\begin{center}
\label{table:charges}
\begin{tabular}{|c|c|c|c|c|}
\hline
~ 	&$x^i$			&$\p\xb^i$				&$\gamma^i$			&$\gammab^i$  \\ \hline
$q$	&$\alpha_i$		&$-\alpha_i$			&$\alpha_i-1$			&$1-\alpha_i$	\\ \hline
$2h$	&$\alpha_i$	&$2-\alpha_i$	&$1+\alpha_i$ 	& $1-\alpha_i$ \\ \hline
$\qb$ &$\alpha_i$		&$-\alpha_i$			&$\alpha_i$			&$-\alpha_i$ \\ \hline
\end{tabular}
\caption{Weights and charges of the fields}
\end{center}
\end{table}

The left-moving $N=2$ algebra in $\Qb$ cohomology is perfectly 
suited to study supersymmetric ground states of the theory.  In
string theory, these right-moving Ramond sector states correspond
to massless fermions in the space-time theory, and in a space-time
supersymmetric compactification knowledge of these states is
sufficient to reconstruct the massless spectrum.  While in general
determining the $\Qb$ cohomology is a challenge, the 
super-renormalizability of the \LG\ theory reduces the computation
to two simple steps~\cite{Kachru:1993pg}:  the UV theory is truncated
to the fields in the $N=2$ algebra of (\ref{eq:Tj}), and the $\Qb$ operator acts on the remaining modes via
\begin{equation}
\Qb = \oint \frac{dz}{2\pi i} \gamma^i \p_i W(x).
\end{equation}

\subsection{The \LG\ Orbifold} \label{ss:lgorb}
In order to apply these ideas to string compactifications, they must
be generalized to \LG\ orbifolds.  This has
been carried out in~\cite{Kachru:1993pg} and extended to a wide
class of (0,2) heterotic backgrounds in~
\cite{Distler:1993mk,Kawai:1994qy,Distler:1995mi}.   
In this section we will review the (2,2) heterotic models.  

Following Gepner~\cite{Gepner:1987qi}, we know that a $c=\cb = 9$ 
(2,2) SCFT with integral $q-\qb$ charges can be used to construct a
space-time supersymmetric $\GE_6\times\GE_8$ heterotic
compactification.  In \LG\ models the natural
way to achieve this is to orbifold by $\exp(2\pi i J)$~\cite{Vafa:1989xc}.
In the Gepner construction, the ``internal'' theory is tensored with a free
theory of $10$ left-moving Majorana-Weyl fermions $\lambda^A$, as well 
as a level-one left-moving $\GE_8$ current algebra.  Adding four (0,1)
superfields for the four Minkowski directions leads to correct central
charges for a critical heterotic string, and modular invariance and 
space-time supersymmetry are ensured by performing a GSO projection
on the world-sheet fermion numbers on the left and the right.  The 
right-moving GSO projection is, as usual, responsible for space-time
supersymmetry, while the left-moving projection ensures that the manifest
$\SO(10)\times\GU(1)$ gauge symmetry is enhanced to 
$\GE_6$.\footnote{The $\GU(1)$ factor is precisely the left-moving 
R-symmetry.} 

In the (2,2) \LG\ theory, the GSO projections are conveniently combined
with the orbifold onto integral $q$:  we simply orbifold by $g=\exp(-i\pi J)$
and project onto suitable fermion numbers, both on the left and right,
mod $2$.  The massless fermions that are $\GE_6\times\GE_8$ singlets
come from the (NS, R) sectors, i.e. the sectors twisted by $g^k$ with $k$
odd.  In such a sector the GSO projection amounts to keeping states with
integral $q$.  Since $\alpha_i$ are rational, $g^{2d} =1$ for some integer
$d$, and hence there are $2d-1$ twisted sectors.  Space-time CPT
exchanges the $k$-th twisted sector with the $(2d-k)$-th sector, so we
need only consider $k=1,3,\ldots,d$.

Massless fermion states must be in the $\Qb$ cohomology, and
level matching implies that the left-moving energy $E$ must vanish.  The
$\GE_6$ representation, as well as the type of space-time multiplet
(vector or chiral/anti-chiral) is determined by the $q,\qb$ charges.  The 
$\GE_6$ representation follows from the standard decomposition
\begin{align}
\GE_6 &\mapsto \SO(10) \times \GU(1) \nonumber \\
\rep{78} &\mapsto \rep{45}_0 + \rep{16}_{-3/2} + \brep{16}_{3/2} + \rep{1}_0 \nonumber\\
\rep{27} &\mapsto \rep{16}_{1/2} + \rep{10}_{-1} + \rep{1}_{2} \nonumber \\
\brep{27} &\mapsto \brep{16}_{-1/2} + \rep{10}_{1} + \rep{1}_{-2}.
\end{align}
The $\GE_6$ singlet states must have $q = 0$.  The type of space-time multiplet
can be determined by working out the action of spectral flow on
corresponding (NS,NS) operators.  The result is that a massless fermion
with right-moving charge $\qb = \pm \ff{3}{2}$ is a vector; if $\qb = -\ff{1}{2}$ it 
belongs to a chiral multiplet, and if $\qb = \ff{1}{2}$, it belongs to an anti-chiral
multiplet.  

The algorithm for determining the singlet spectrum is therefore quite simple:
for each odd sector in the \LG\ orbifold, we must identify states in the
$\Qb$ cohomology with $E=q=0$ and $\qb = \pm \ff{1}{2}$.  Since we restrict
to the $\Qb$ cohomology the left-moving quantum numbers can be
determined from the left-moving $N=2$ UV algebra.  It is useful to note
that $\Qb$ commutes with the left-moving algebra and has a definite
right-moving charge $\qb = 1$.  Thus, in any sector the zero energy
states live in a complex
\begin{equation}
\xymatrix{\cdots  \ar[r]  &U_{-3/2} \ar[r]^-{\Qb} &  U_{-1/2} \ar[r]^-{\Qb}  &  U_{1/2} \ar[r]^-{\Qb} &  U_{3/2} \ar[r] & \cdots,}
\end{equation}
where the $U_{\qb}$ are states with definite $\qb$ charge.

\subsection{Quantum Numbers in Twisted Sectors} \label{ss:qtwist}
The remaining question to be addressed is the computation of the $E$, $q$ 
and $\qb$ quantum  numbers in the twisted sectors.  The $E$ and $q$ quantum
numbers of the twisted vacua can be obtained by using the UV $N=2$ algebra.
In the $k$-th sector the fields have a twisted moding 
\begin{align}
x^i(z) &= \sum_{s\in \Z-\nu_i} x^i_s z^{-s-h_i},      
&\qquad
& \gamma^i(z)  = \sum_{s\in \Z-\nut_i} \gamma^i_s z^{-s-\htld_i} \nonumber\\
2\p\xb^i(z) &= \sum_{s\in\Z+\nu_i} \rho^i_s z^{-s+h_i-1}, 
&\qquad 
&\gammab^i(z) = \sum_{s\in\Z+\nut_i} \gammab^i_s z^{-s+\htld_i-1},
\end{align}
where $2h_i = \alpha_i$, $2\htld_i = 1+\alpha_i$, and
\begin{alignat}{2}
\label{eq:nus}
\nu_i  &= \frac{k\alpha_i}{2}\pmod 1 &\qquad 0&\le\nu_i<1\nonumber\\
\nut_i &= \frac{k(\alpha_i-1)}{2} \pmod 1 &\qquad-1&<\nut_i\le 0\ .
\end{alignat}

The Fock vacuum $|k\ra$ is defined as the state annihilated by all positive modes.
The left-moving quantum numbers of $|k\ra$ can be computed by working out 
the one-point functions of $J$ and $T$ via the mode expansion.  For instance, we
easily find
\begin{align}
2\la k| x^i(w+\ep) \p\xb^i(w) |k\ra &= \ep^{-1} + (\nu_i-h_i) w^{-1}+ O(\ep), \nonumber\\
\la k| \gamma^i(w+\ep) \gammab^i(w) |k\ra &= \ep^{-1} + (1+\nut_i-\htld_i) w^{-1}+ O(\ep).
\end{align}
Using a point-splitting regularization for $J$, these correlators imply
\begin{align}
w\la k| J(w)  |k\ra = \sum_i \left[ (\alpha_i-1) (1+\nut_i-\htld_i) 
-\alpha_i (\nu_i -h_i)\right],
\end{align}
whence we read off the left-moving charge of $|k\ra$ as
\begin{equation}
q_k = \sum_i \left[ (\alpha_i-1) (\nut_i+\ff{1}{2}) -\alpha_i (\nu_i -\ff{1}{2})\right].
\end{equation}
Similarly, we obtain the weight via $w^2 \la k |T(w)|k\ra = h_k$:
\begin{equation}
h_k =  \frac{3}{8} + \Delta_k, \qquad \Delta_k =\frac{1}{2} \sum_i \left[ \nu_i(1-\nu_i) +\nut_i(1+\nut_i) \right].
\end{equation}
To obtain the energy, we must transform from the plane to 
the cylinder and remember to include contributions from the other left-moving
degrees of freedom.  In sectors with $k$ even, we find $E_k= 0$---an answer
in line with the (2,2) supersymmetry.  For $k$ odd, the
result is
\begin{equation}
E_k = \underbrace{\frac{3}{8} +\Delta_k - \frac{9}{24}}_{\text{LG}} + \underbrace{0 - \frac{5}{24}}_{\SO(10)} + \underbrace{0 -\frac{8}{24}}_{\GE_8, ~\text{NS}}+\underbrace{0 - \frac{2}{24}}_{\text{space-time}} = -\frac{5}{8} + \Delta_k.
\end{equation}

The remaining quantum number is the right-moving R-charge.  The simplest
way to determine this is to note that in the UV the right-moving current 
$\Jb$ is related to the left-moving current via $\Jb = J + J_B$, where
$J_B$ assigns charge $+1$ to $\gamma^i$ and $-1$ to $\psi^i$.  The 
$J_B$ symmetry is independent of $W$, and we can evaluate the charge
of the twisted vacuum by simply setting $W=0$.  Combining this with
the result for $J$, we find
\begin{equation}
\qb_k = \sum_i \left[ \alpha_i (\nut_i + \ff{1}{2}) + (\alpha_i -1) (-\nu_i + \ff{1}{2}) \right].
\end{equation}

Having determined the ground state quantum numbers, we will construct $E=q=0$
energy states by acting on $|k\ra$ with the lowest excited modes, which we will denote
by 
\begin{equation}
x_i \equiv x^i_{-\nu_i}, \quad \rho_i \equiv \rho^i_{\nu_i-1}, 
\quad \gamma_i \equiv \gamma^i_{-1-\nut_i}, \quad
\gammab_i \equiv \gammab^i_{\nut_i}.
\end{equation}
As usual, an oscillator with mode $\nu$ will contribute $-\nu$ to 
the energy.
In describing the $\Qb$ cohomology we will also need the conjugate modes, which we 
will distinguish with a dagger:
\begin{equation}
x_i^\dag \equiv \rho^i_{\nu_i}, \quad \rho_i^\dag \equiv x^i_{1-\nu_i}, \quad
\gamma^{\dag}_i \equiv \gammab^i_{1+\nut_i}, \quad
\gammab^{\dag}_i \equiv \gamma^i_{-\nut_i}.
\end{equation}
The $\Qb$ operator takes the general form
\begin{equation}\label{eq:LGqbar}
\Qb = \sum_i \left\{ \gamma_i \p_iW_{1+\nut_i} + \gammab_i^{\dag} \p_iW_{\nut_i}\right\}.
\end{equation}

\subsection{The Quintic} \label{ss:lgquinitc}
We will now apply the method reviewed above to determine the singlet 
spectrum at the \LG\ point in the moduli space of the quintic in $\C\P^4$.  
This calculation was performed in \cite{Kachru:1993pg}, and we repeat it
here to review the notation and also to remark on some universal features
of such models.
The theory is described by the superpotential $W(X_0,\ldots, X_4)$, a degree
five polynomial. The R-charges of the  fields are $\alpha_i=\ff{1}{5}$, so that the
$\Z_5$ orbifold acts by multiplying the fields by fifth roots of unity.  Thus, there
are ten twisted sectors.  CPT invariance allows us to restrict to $k =0, \ldots, 5$,
and since we are interested in the singlets, we can restrict to
odd $k$.  The  ground state quantum numbers and field modings for the sectors
are  given in table~\ref{table:quintic}.

\begin{table}[t]
\renewcommand{\arraystretch}{1.3}
\begin{center}
\begin{tabular}{|c|c|c|c|c|c|c|}
\hline
$k$ 		&$E_{k}$			&$q_{k}$		&$\qb_{k}$ 	&$\nu$		&$\nut$			\\ \hline
$0$		&$0$				&$-\ff{3}{2}$	&$-\ff{3}{2}$	&$0$			&$0$				\\ \hline
$1$		&$-1$			&$0$			&$-\ff{3}{2}$	&$\ff{1}{10}$	&$-\ff{2}{5}$		\\ \hline
$2$		&$0$				&$\ff{3}{2}$	&$-\ff{3}{2}$	&$\ff{1}{5}$	&$-\ff{4}{5}$		\\ \hline
$3$		&$-\half$			&$-1$		&$-\ff{1}{2}$	&$\ff{3}{10}$	&$-\ff{1}{5}$		\\ \hline
$4$		&$0$				&$\half$		&$-\half$		&$\ff{2}{5}$	&$-\ff{3}{5}$		\\ \hline
$5$		&$0$				&$-2$		&$\ff{1}{2}$	&$\ff{1}{2}$	&$0$				\\ \hline
\end{tabular}
\caption{Ground state quantum numbers and modings for the quintic.}
\label{table:quintic}
\end{center}
\end{table}

Zero energy states with $q= 0$ are only found in the $k=1$ and $k=3$ sectors.  
An explicit basis for these states is 
\begin{equation}
\xymatrix@C=40mm@R=5mm{
  \bar q=-\ff32&\bar q=-\ff12\\
{\begin{matrix}
\gamma_i\gammab_j |1\ra_{25}\\\oplus\\
x_i \rho_j |1\ra_{25}\end{matrix}}
\ar[r]^-\Qb&
F_{[4]}^i \gamma_i |1\ra_{350}\\
&x_i \gammab_j |3\ra_{25}
} \label{eq:quinLG}
\end{equation}
Unless indicated otherwise, here and in what follows $F_{[d]}$ denotes
a degree $d$ polynomial in the $x_i$ with respect to the multi-grading
of the relevant homogeneous coordinate ring (in this case the
$x_i$ just have charge $1$). The subscript of the ket indicates the
number of linearly independent states of each type.  The map $\Qb:
U_{-3/2} \to U_{-1/2}$ is given by
\begin{equation}
\Qb = \gammab^{\dag}_i \p_i W + \gamma_i \p_{ij} W \rho_j^\dag,
\end{equation}
so that the image of an arbitrary state $|\psi\ra \in U_{-3/2}$, specified by two 
matrices $c^{ij}$ and $d^{ij}$ is
\begin{align}
\Qb |\psi \ra &= \Qb \left\{4 c^{ij} \gamma_i \gammab_j + d^{ij}  x_i \rho_j \right\}|1\ra 
= \left\{ -4c^{ij}\p_j W + d^{kj}  x_k \p_{ij} W \right\} \gamma_i |1\ra.
\end{align}
What is the kernel of $\Qb$?  Since $W$ is quasi-homogeneous, $\dim
\ker \Qb \ge 1$ for any $W$.  A $\Qb$-closed state is obtained by
setting $c^{ij}= d^{ij} = \delta^{ij}$.  In fact, for generic $W$ this
is the only $\Qb$-closed state in $U_{-3/2}$.  Its existence is not
surprising: this is the vector multiplet corresponding to the unbroken
$\GU(1)_L$ symmetry.  Thus, we find that there are generically $301$
massless chiral singlets in the $k=1$ sector.

When $W$ is taken to be Fermat, we find that there are five
$\Qb$-closed states at $U_{-3/2}$, with $d^{ij} = d^i \delta^{ij}$ and
$c^{ij} = d^{ij}$.  Again, this is not a big surprise, since at this
point in the moduli space the theory reduces to the Gepner model for
$\oplus_{i=1}^5 A^i_{4}$.  This theory has five unbroken $\GU(1)$
currents, each of which leads to a massless vector multiplet.  So, at
the Fermat point we find $305$ massless chiral singlets at $k=1$.

The physics encoded by the change in $\ker \Qb$ is just the
supersymmetric Higgs mechanism: the disappearance of a massless vector
multiplet is accompanied by the disappearance of a massless $\GE_6$
singlet chiral multiplet.  The number of massless vector multiplets is
given by the number of decoupled components of the \LG\ theory.  For
instance, turning on the unique monomial containing all the fields
breaks all but one of the currents and leads to $301$ massless
singlets.

Finally, we turn to the $k=3$ sector.  Here there are $25$ zero energy states
with $q=0$ as shown in (\ref{eq:quinLG}).
Clearly $\Qb = 0$ in this sector, so that all of these states correspond to
massless singlets.  Combining these states with the $k=1$ singlets, we
see that the theory contains $326$ massless singlets for generic $W$
and $330$ at the Fermat point.  Subtracting the $1+101$ (2,2) moduli,
we find $224$ singlets that are not associated to extra $\GU(1)$ gauge
symmetries.

The (2,2) \LG\ Lagrangian can be deformed to a (0,2) theory by 
replacing $\p_i W$ with arbitrary quartic polynomials $W^i$ in the 
component Lagrangian.  Although the
$G^\pm$ generators of the left-moving SCA are no longer conserved,
the $J$ and $T$ still generate a left-moving $\GUL \times$ Virasoro
algebra.  The computation of the ground-state quantum numbers and
$\Qb$ cohomology are unchanged by these deformations.  

\subsection{Universal Structure in LG Singlet Spectrum}\label{ss:lguni}

The spectrum of singlets we have seen in the quintic in fact exhibits
a pattern that is somewhat universal and will be useful in connecting
the \LG\ results to calculations in the large radius phase.
Consider a \LG\ model with $N=5$ fields and arbitrary charges
$0<\alpha_i\le \ff{1}{2}$ such that $\sum_i(1-2\alpha_i)=3$.  We will
consider the $k=1$ and $k=3$ sectors of any such model and find that
some singlet states are universally present.

To simplify expressions, note from (\ref{eq:nus}) we have
\begin{equation}
\nut_i =
\begin{cases}
\nu_i-\ff{1}{2}, & 0<\nu_i\le \ff{1}{2}\\
\nu_i-\ff{3}{2}, &  \ff{1}{2}\le\nu_i<1.
\end{cases}
\end{equation}

We begin with the $k=1$ sector.  Here 
\begin{align*}
0 < \nu_i = \ff12\alpha_i \le
\ff{1}{4}
\quad\text{and}\quad
-\ff{1}{2}<\nut_i = \nu_i-\ff{1}{2}\le-\ff{1}{4}.
\end{align*}
Inserting these values we
find $E_1=-1$ and $(q_{1},\qb_1 = (0,-\ff{3}{2})$.  These values show
immediately that zero-energy states will be given by 
\begin{equation}\label{eq:kone}
|\psi\ra = \left( F(x,\gammab) + G^i(x,\gammab)\rho_i + H^i(x,\gammab)\gamma_i\right)|1\rangle\ .
\end{equation}
Since $\nu_i\propto\alpha_i$ functions of $x$ can be classified by
their degree, $f_{[p]}(\lambda^{\alpha_i d}x_i) = \lambda^p f(x)$, so
that $p$ determines the charge and energy of $f(x)$.   We also use the 
notation $n_i = \alpha_i d\in\Z$.  A possible chiral field with
$\alpha=\ff{1}{2}$ must here be distinguished, so we use the somewhat clumsy
notation 
\begin{align}
&\alpha_I = \ff{1}{2}\qquad &I&=0,\ldots, n-1 \ ;\nonumber\\
0 < &\alpha_a<\ff{1}{2}\qquad&a&=n,\ldots, 4 \ . 
\end{align}
Clearly, $0\le n\le 1$; if $n=0$ the first row is vacuous.

Considering the first term in (\ref{eq:kone}), the zero energy
condition shows that
\begin{align}
  F(x,\gammab) &= F_{[2d]}(x) + F^i_{[d+n_i]}(x)\gammab_i + 
F^{ij}_{[n_i+n_j]}(x)\gammab_i\gammab_j
  + \cdots \nonumber\\
  &\qquad+ F^{i_1\ldots i_m}_{[n_1 + \ldots + n_m-(m-2)d]}
\gammab_{i_1}\cdots\gammab_{i_m} + \cdots.
\end{align}
The expansion extends so long as there are terms for which the degree of $F$ is positive.  All of these yield states at $q=2$.
The second term potentially yields two types of zero energy states
\begin{align}
G^I(x,\gammab) &= G_{[n_I]}^I(x) + \widetilde G^I\gammab_I\nonumber\\
G^a(x,\gammab) &= G_{[n_a]}^a(x)\ .
\end{align}
States of the first type have $(q, \qb) = (0,-\ff{3}{2})$ and represent
singlets in vector multiplets.  The second term in the first line
contributes $n$ states at $(q,\qb) = (0,-\ff{5}{2})$.
The third term also yields two types of zero energy states
\begin{equation}
H^i(x,\gammab) = H^i_{[d-n_i]}(x) + H^{ij}_{[n_j-n_i]}(x)\gammab_j\ .
\end{equation}
The first of these contributes states at $(q,\qb) = (0,-\ff{1}{2})$
(chiral matter) and the second at $(0,-\ff{3}{2})$ (vector multiplets).
This exhausts the possible singlet states at $k=1$.  The action of
$\Qb$ is determined from (\ref{eq:LGqbar}) 
\begin{align}
\Qb(\widetilde G^I\gammab_I \rho_I)= \widetilde G^I \p_I W\rho_I, \quad
\Qb(G^i \rho_i) = \sum_kG^i \p_{ik} W\gamma_k, \quad
\Qb(H^{ij} \gammab_j\gamma_i )= H^{ij}\p_jW\gamma_i\ .
\end{align}

We now move to the $k=3$ sector.  In general we will not enumerate all
possible states at $q=E=0$ or the action of $\Qb$.  Rather, we will
show that certain states arise universally at $\qb=-\ff{1}{2}$.  Here a
different distinction among the fields by weight is appropriate
\begin{align}
\ff{1}{3} <&\alpha_A\le \ff{1}{2}&\ff{1}{2}<&\nu_A\le \ff{3}{4}  &-1<&\nut_A\le-\ff{3}{4}&A&=0,\ldots, m-1\ ;\nonumber\\
0< &\alpha_a\le \ff{1}{3} &0<&\nu_a\le\ff{1}{2}  &-\ff{1}{2}<&\nut_a\le0&a&=m,\ldots,4 \ .
\end{align}
Clearly $n\le m\le 2$.
Inserting these values we find that the ground state is characterized
by 
\begin{align}
2E_3 & =  m-1-3\sum_A\alpha_A; \nonumber\\
(q_{3},\qb_{3}) &=
 \left(m-1-\sum_A\alpha_A,\ -\ff{1}{2}-\sum\alpha_A\right).
\end{align}
We will not attempt a complete characterization of all zero-energy
states in this sector but note that the following states always arise:
\begin{equation}\label{eq:uni3}
  K^A_{[n_A]}\prod_{B\neq A}\gamma_B |3\ra,\quad
  K^a_{[n_a]}(x)\gammab_a\prod_A\gamma_A
  |3\ra.
\end{equation}
Note that since we still have $\nu_i\propto\alpha_i$ we classify
functions of $x$ by degree as above.  All of these states have
$(q,\qb) = (0,-\ff{1}{2})$ and represent scalars in chiral multiplets.  The
first term obviously contributes when $m$ is nonzero.  The coefficient
space for these states is identical to the first row of $\qb=-\ff{3}{2}$ states
in the $k=1$ sector  in table \ref{t:LGuni} and in terms of counting
singlets they explicitly ``cancel them out.''  In
general there can be other zero energy states at $k=3$.  These can
include states with charge $(0,-\ff{3}{2})$ and nontrivial $\Qb$ action, so
not all of the states listed above are physical.  In all cases we have
observed, the coefficient spaces of these $\qb=-\ff{3}{2}$ states are then
repeated as coefficients of $\qb=-\ff{1}{2}$ states in other sectors,
effectively canceling again.  We term this reappearance of $k=1$
states in higher sectors the ``cascade.''  We will see an example of
this in Section \ref{ss:LG7}.

\begin{table}[!t]
\vspace{-10mm}
\[
\xymatrix@C=13mm@R=5mm{
  \bar q=-\ff52&\bar q=-\ff32&\bar q=-\ff12\\
{\begin{matrix}
\widetilde G^I_{[0]}\gammab_I\rho_I|1\rangle \ar[r]^-\Qb
\end{matrix}}&
{\begin{matrix}
\oplus_i G^i_{[n_i]}\rho_i|1\rangle \\\oplus\\
\oplus_{n_j\ge n_i} H^{ij}_{[n_j-n_i]}\gammab_j\gamma_i|1\rangle 
\end{matrix}}
\ar[r]^-\Qb&
\oplus_k H^k_{[d-n_k]}\gamma_k|1\rangle \\
&&{\begin{matrix}
\oplus_A K^A_{[n_A]}\prod_{B\ne A}\gamma_B|3\ra\\\oplus\\
\oplus_a K^a_{[n_a]}\gammab_a\prod_A\gamma_A|3\rangle
\end{matrix}} 
}
\]
\caption{Universal Singlets in \LG\ models.}  \label{t:LGuni}
\end{table}

The set of states we have listed here is present in any \LG\ model where
the orbifold simply projects onto integral R-charges.
In the case of the quintic this is the complete complement of
singlets. In other models there will be additional states in various
sectors, but these universally present states will figure in comparing
the \LG\ phase with the orbifold.

\subsection{More LG Orbifolds} \label{ss:lgorba}

The methods described above are easily extended to study quotients
of the \LG\ models discussed above by further discrete symmetries.
These will be useful when constructing the mirrors as \LG\ 
models.\footnote{In addition, a generic \LG\ phase of a GLSM 
will have such further quotients as part of the discrete gauge symmetry.}
A model with charges $\alpha_i = n_i/d$ has a natural non-$R$ action
of $\GU(1)^N$ acting by phases on the worldsheet chiral multiplets
$\Phi_i$.  This is broken by the superpotential $W$, and for generic
$W$ it is broken to $\Z_d$, by which we took the quotient above.
Special nonsingular superpotentials leave a larger discrete subgroup
$G \subset \GU(1)^N$ unbroken, and we can construct quotient theories 
following the
same steps as above, with the simplifying feature that the new
symmetries act non-chirally.  In what follows we will take $G$
to be abelian, and we will not consider general choices of
discrete torsion in \LG\ orbifolds~\cite{Intriligator:1990ua}.

The symmetry groups by which we quotient will act by phases on 
the chiral multiplets.  We represent the group elements by $N$-tuples
of rational numbers defined up to integers, so that the vector $w$
represents the action of $g \in G$:
\begin{equation}\label{eq:lgorb}
g : \Phi_i\mapsto e^{2\pi i w_i}\Phi_i\ ,
\end{equation}
generating a cyclic action of order $d(w) = \gcd(w_i)$. 

To construct the quotient by a group generated by vectors $w^{(a)}$ we
introduce twisted sectors labeled by $0\le t^{(a)}\le d(w^{(a)})-1$.
The new symmetry is not an $R$-symmetry so the twists of bosons and
fermions (\ref{eq:nus}) are modified to
\begin{align}
\nu_i &= \frac{k\alpha_i}2 + \sum_a t^{(a)} w^{(a)}_i\pmod 1
&0\le\nu_i<1\nonumber\\
\nut_i &= \frac{k(\alpha_i-1)}2 +  \sum_a t^{(a)} w^{(a)}_i\pmod 1
&-1< \nut_i\le 0\ .
\end{align}
We also introduce a projection onto states invariant under
(\ref{eq:lgorb}).  

The $G$ action in twisted sectors is determined by the requirement of
modular invariance from (\ref{eq:lgorb}).  For a non-chiral symmetry
this is the same as the action in the untwisted sector, and the
twisted vacua are uncharged.  The GSO projection, however, is chiral
and thus sectors with nonzero $k$ will in general carry 
a $G$-charge.\footnote{We would like to thank B.~Wurm for his help in
clarifying this point.}  The charge is easily computed by working with the
UV fields and free OPEs.  Since the twisted bosons make no contribution, 
it is sufficient
to consider the action of $g$ on the fermions, which we express as a
subgroup of the $\GU(1)$ vectorial symmetry with current
$J_G = \sum_i w_i (\gamma^i\gammab^i+\psi^i\psib^i).$
The  full $\GU(1)$ symmetry is broken by the superpotential couplings, but
since $W$ is $G$-invariant, we can use this embedding to compute the
$G$-charge.  The result is that a twisted vacuum  $|k, t^{(a)}\ra$ transforms
by a phase $e^{2\pi i q_g}$, with 
\begin{align}
\label{eq:orbcharge}
q_g = \sum_i w_i (\nut_i -\nu_i +1) \pmod 1\ .
\end{align}

The quotients of interest to us will be those preserving space-time
supersymmetry.  This requires that the left-moving spectral flow
operator, in the $k=1$ sector, be preserved by the projection.  
 We can construct  this
operator in the free field representation of section~\ref{ss:lggen}
and find that it is preserved if 
\begin{equation}\label{eq:wSUSY}
\sum_i w_i = 0\ .
\end{equation}
Note that since (\ref{eq:lgorb}) depends on $w_i$ only modulo
integers, this condition is equivalent to the more familiar 
\begin{equation}
\label{eq:wSUSYii}
\sum_i w_i\in \Z\ 
\end{equation}
in terms of restricting the allowable quotients.  In the twisted
sectors, (\ref{eq:orbcharge}) will hold when $w$ are chosen to satisfy
the more stringent condition.

\subsection{Gepner Models and Mirror Symmetry}\label{ss:Gepner}

For special values of the superpotential couplings, the \LG\ model in
all of our examples is an exactly solvable theory \cite{Gep:}.  Prior to the
orbifold of section \ref{ss:lgorb} we have a product of (2,2) minimal models.
This, of course, allows a calculation of the spectrum of singlets at
this point in the moduli space, but this is equivalent to a special
case of  the \LG\ calculation, as discussed above.  The utility, to
our work, of the Gepner model, is that at this point we have
a construction \cite{GP:orb}  of the mirror model as an orbifold, as
well as an explicit mirror map in terms of the Gepner construction.
This allows us to find the singlet states in the mirror model
corresponding to the states we enumerate using the \LG\ construction.
We can then construct the mirror as a \LG\ orbifold and study the
dependence of the singlet spectrum on the mirror superpotential.
Since deformations of the mirror superpotential are mapped to K\"ahler
deformations in the original model, we can thus predict which singlet
states will be lifted by K\"ahler deformations away from the \LG\
locus.

In all of our examples, the exactly solvable superpotential will
be a sum of terms of the
form $x_i^{k_i+2}$, corresponding to an $A_{k_i+1}$ minimal model at
level $k_i$ and of the form $x_i^{l_i+1} + x_i y_i^2$, corresponding to
a $D_{l_i+2}$ minimal model at even level $k_i=2l_i$, leading to a 
tensor product of $n$ minimal models.\footnote{We hope there will 
be no confusion between the minimal model levels labeled by $k_i$ 
and the twisted sectors labeled by $k$.}  Primary fields in the level-$k$
minimal model are labeled $\Phi^{l,\bar l}_{q,s;\bar q,\bar s}$ where
$0\le l,\bar l \le k$, subject to the identifications
\begin{equation}
q \sim  q + 2(k+2)\qquad s \sim  s + 4\ ,
\end{equation}
(and the same for $\bar q, \bar s$) as well as 
\begin{equation}\label{eq:minid}
\Phi^{l,\bar l}_{q,s;\bar q,\bar s} \sim \Phi^{k-l,k-\bar
  l}_{q+k+2,s+2;\bar q+k+2,\bar s+2}\ .
\end{equation}
Fields with even (odd) $s$ create states in the NS (R)
sector.\footnote{In fact, $s=2$ states are not primary but after the
  orbifold they do create highest weight states; this is related to
  the fact that the quotient projects out some modes of the
  supercurrents in the individual minimal models.}  We
use Gepner's notation $\lqs{l}{q}{s}|\lqs{\bar l}{\bar q}{\bar s}$ for
the state created by (\ref{eq:minid}).  The 
$R$-charge and conformal weight of a state are given by 
\begin{align}
q & =  -\sum_i \frac{\tilde q_i}{ k_i+2} + \frac{s_i}{2}\nonumber\\
h &= \sum_i \frac{l_i(l_i+2)-q_i^2}{4k_i+8}+\frac{s_i^2}{ 8}\ ,\\
\end{align}
for $q_i,s_i$ in the standard range $|q_i-s_i|\le l$, $-1\le s_i\le
2$, where $\tilde q_i = q_i +s_i$ for R states and $\tilde q_i = q_i$
for NS states.

The minimal model at level $k$ has a partition function
\begin{equation}
Z = \frac{1}{ 2}\sum_{l+q+s  \in 2 \Z} A_{l,\bar
  l}\chi^l_{q,s}\chi^{\bar l\,*}_{q,s}
\end{equation}
where $A_{l,\bar l}$ is the appropriate affine modular invariant at
level $k+2$ and the factor of $\ff{1}{2}$ reflects the identification
(\ref{eq:minid}). The model enjoys a discrete symmetry $G_k =
\Z_{k+2}\times\Z_2$ under which the state $\lqs{l}{q}{s}$ has weights
$q,s$.  In the associated \LG\ model we will be interested in the
$\Z_{k+2}$ subgroup of this generated by $x_i\mapsto e^{2\pi
  i/k_i+2}x_i $ in the $A_{k_i+1}$ model and by $x_i\mapsto
e^{2\pi i/l_i+1}x_i $ and $y_i\mapsto -y_i$ in the $D_{l_i+2}$ model.

The Gepner construction of a string vacuum as a quotient of the tensor
product was introduced in Section \ref{ss:lgorb}.  We
add free fields and perform a quotient by $\Z_d\times \Z_2^{n}$.
The quotient introduces twisted sectors in which $\bar q_i,\bar s_i$
differ from $q_i,s_i$ by $k$ and additional twists in which any two
$\bar s$ indices are shifted by 2.  The gauge symmetry of the model is
$\GE_6\times \GE_8\times \GU(1)^{{n}-1}$.  The $\SO(10)$-neutral
scalars in chiral multiplets are states with
$\bar q=1, \bar h=\ff{1}{2}$ and $q = h = 0$.  The corresponding fermion
states are obtained by applying the spacetime supersymmetry
generator, shifting $\bar q,\bar s$ by one, and leading to $\bar
q=-\ff{1}{2}$. We will denote states in the resulting model by
\begin{equation*}
\lqs{l_1}{q_1}{s_1} \cdots \lqs{l_{n}}{q_{n}}{s_{n}} \Big|
\lqs{\bar l_1}{\bar q_1}{\bar s_1} \cdots\lqs{\bar l_{n}}{\bar
 q_{n}}{\bar s_{n}} .
\end{equation*}

The mirror model is constructed as a further quotient by a subgroup of
$G$, essentially the subgroup under which the spacetime supercharge is
invariant.  The quotient introduces twisted sectors in which $\bar
q_j$ is shifted relative to $q_j$ by $2t_a m^{(a)}$ for a lattice
generated by a set of integer vectors $m^{(a)}$ (and the associated
projection).  The result of the construction is \cite{GP:orb} a model
in which the primary fields are related to those of the original
theory by $q\mapsto -q,\ s\mapsto -s$.  The orbifold construction can
be realized in the \LG\ model as a quotient following Section
\ref{ss:lgorba} with the action on the chiral superfields given by
\begin{align*}
w^{(a)}_j = \frac{m^{(a)}_j}{k_j+2} \qquad\qquad\qquad & \text{for $A$-type $x_j^{k_j+2}$}\\
w^{(a)}_{j,x} = \frac{2m^{(a)}_j}{ k_j+2}, \qquad
w^{(a)}_{j,y} = \frac{m^{(a)}_j}{ 2} \qquad
&\text{for $D$-type $x_j^{(k_j+2)/2} + x_j y_j^2$}
\end{align*}

To use mirror symmetry to study the behavior of the singlet spectrum
under K\"ahler deformations away from the Gepner point, we first find
the spectrum of singlets in the original model.  For each of these we
construct the mirror state and identify the twisted sector in the
mirror quotient in which it arises.  We then consider the mirror \LG\
model constructed as an orbifold.  The mirror superpotential will
admit polynomial deformations related by the monomial-divisor mirror
map to the toric K\"ahler deformations in the original model.  In the
relevant twisted sectors, we can study the change in $\Qb$ cohomology
when the superpotential is deformed, thus identifying which singlet
states are lifted under K\"ahler deformations away from the Gepner
point.


\section{Deformations of the Tangent Bundle} \label{s:T}

\subsection{The tangent sheaf} \label{ss:tan}

Let $T$ denote the tangent sheaf (or bundle --- we will use the terms
interchangeably here) of a \CY\ threefold $X$. We are interested in
first order deformations of $T$ since they correspond classically to
massless fields allowing, to first order, a deformation of a
(2,2)-model to a (0,2)-model.

Such deformations are given by $\Ext^1_X(T,T)=H^1(X,\End(T))$. For a
simple argument we refer to \cite{GSW:book}, chapter 15. Methods of
computing $H^1(\End(T))$ have been studied for some time
\cite{Eastwood:1989sd,Berglund:1990rk,Hub:book} for cases of \CY\
manifolds in products of projective spaces. Here we give a method
that is reasonably direct for complete intersections in toric
varieties.

Let $V$ be a compact toric variety and let $X$ be a \CY\ complete
intersection within $V$. That is, we are in the context of \CY's as
studied in \cite{Bat:m,BB:mir}.  We quickly review the construction of
$V$ to fix notation.  Let $x_0,\ldots,x_{N-1}$ be the homogeneous
coordinates on $V$. That is, we have a homogeneous coordinate ring in
the sense of Cox \cite{Cox:}
\begin{equation}
  R = \C[x_0,\ldots,x_{N-1}].
\end{equation}

We now have a short exact sequence
\begin{equation}
\xymatrix@1{
  0\ar[r]&M\ar[r]&\Z^{\oplus N}\ar[r]^\Phi& D\ar[r]&0,
} \label{eq:MZD}
\end{equation}
where $D$ is a lattice\footnote{Assumed to be torsion-free.} of rank
$r$. Each column of the matrix $\Phi$ can be thought of as a
$\GU(1)^r$ charge vector of the coordinates $x_i$. That is, $R$ has
the structure of an $r$-multigraded ring.

The toric variety is given as
\begin{equation}
  V = \frac{\Spec(R) - Z(B)}{(\C^*)^r}, \label{eq:V}
\end{equation}
where $B$ is the ``irrelevant ideal'' in $R$ and $Z(B)$ is the
associated subvariety of $\C^N$. $B$ is determined combinatorially
from the fan describing $V$.

Let $\mathbf{v}$ denote an element of the lattice $D$, i.e., an
$r$-vector. If $M$ is a multigraded $R$-module then we may shift
multi-gradings to form $M(\mathbf{v})$ in the usual
way. Correspondingly, if $\O_V$ is the structure sheaf of $V$, then we
may denote by $\O_V(\mathbf{v})$ the twisted sheaf associated to the
module $R(\mathbf{v})$. Line bundles on $V$ correspond to
$\O_V(\mathbf{v})$ for various $\mathbf{v}\in D$. If $V$ is smooth
then every element of $D$ defines a line bundle.

Let $\mathbf{q}_i$ denote the row vectors of the transpose of
$\Phi$. That is, $\mathbf{q}_i$ represents the multi-grading of the
homogeneous coordinate $x_i$. Let $T_V$ be the tangent sheaf of
$V$. Assuming $V$ is smooth, we have the generalization of the Euler
exact sequence for a toric variety \cite{BatCox:toric}
\begin{equation}
\xymatrix@1{
  0\ar[r]& \O_V^{\oplus r}\ar[r]^-{x_i\mathbf{q}_i}&
  \displaystyle{\bigoplus_{i=0}^{N-1}
  \O_V(\mathbf{q}_i)} \ar[r]& T_V\ar[r]&0.
} \label{eq:euler}
\end{equation}

Let
\begin{equation}
  \mathbf{Q} = \sum_{i=0}^{N-1} \mathbf{q}_i.
\end{equation}
Suppose $X$ is a smooth hypersurface in $V$ representing the
anticanonical class. Then $X$ is a \CY\ manifold and we have
the adjunction exact sequence:
\begin{equation}
\xymatrix@1{
  0\ar[r]&T_X\ar[r]&T_{V|X}\ar[r]&\O_X(\mathbf{Q})\ar[r]&0,
} \label{eq:adj}
\end{equation}
where we denote a restriction of $\O_V(\mathbf{v})$ to $X$ by
$\O_X(\mathbf{v})$.

Since all the sheaves in (\ref{eq:euler}) are locally-free, we may
restrict to $X$ and the sequence will remain exact. Combining this
with the sequence (\ref{eq:adj}) yields the following fact. The
complex
\begin{equation}
\xymatrix@1{
  0\ar[r]& \O_X^{\oplus r}\ar[r]^-{x_i\mathbf{q}_i}&\displaystyle{
  \bigoplus_i^{\phantom{N}}
  \O_X(\mathbf{q}_i)} \ar[r]^-{\partial_i W}& \O_X(\mathbf{Q})\ar[r]&0,
} \label{eq:TX}
\end{equation}
is exact everywhere except the middle term where the cohomology is
isomorphic to the tangent sheaf $T_X$. Here $W$ denotes the defining
equation for the hypersurface $X$.

It is easy to generalize this to the case of a complete intersection. Suppose
$X$ is defined by an intersection of $W_1=W_2=\ldots=0$. Let each
$W_a$ have grade $\mathbf{Q}_a$. Then the tangent sheaf is given by
the cohomology of
\begin{equation}
\xymatrix@1{
  0\ar[r]& \O_X^{\oplus r}\ar[r]^-{x_i\mathbf{q}_i}&\displaystyle{\bigoplus_i^{\phantom{N}}
  \O_X(\mathbf{q}_i)} \ar[r]^-{\partial_iW_a}&\displaystyle{\bigoplus_a^{\phantom{N}} 
  \O_X(\mathbf{Q}_a)}\ar[r]&0,
} \label{eq:TXci}
\end{equation}
For the remainder of the paper we assume $X$ is a hypersurface.

Before heading into the more complicated $H^1(\End(T_X))$ computation
it will be useful to consider the cohomology of the tangent sheaf itself.
It is most convenient to use the language of the derived category
$\DC(X)$ to manipulate the tangent sheaf. The tangent sheaf is
equivalent in $\DC(X)$ to the complex (\ref{eq:TX}), where the middle
position of (\ref{eq:TX}) is considered position zero. Suppose we have
an object $\cE^\bullet$ in $\DC(X)$ represented by a complex
\begin{equation}
\xymatrix@1{\cdots\ar[r]&\cE^{-1}\ar[r]&\cE^{0}\ar[r]&\cE^{1}\ar[r]&\cdots}
\end{equation}
We can consider the total cohomology (or {\em hypercohomology\/}) of
this complex $H^n(\cE^\bullet)$. There is a spectral sequence
with \cite{CarEil:}
\begin{equation}
  E_1^{p,q} = H^q(\cE^p),
\end{equation}
which converges to $H^{p+q}(\cE^\bullet)$. This gives a method of
computing the cohomology of the tangent sheaf. The $E_1$ stage of the
spectral sequence is
\begin{equation}
\begin{xy}
\xymatrix@C=20mm{
  \vdots&\vdots&\vdots&\\
  H^2(\O_X)^{\oplus r}\ar[r]&\displaystyle{\bigoplus_i^{\phantom{N}}
  H^2(\O_X(\mathbf{q}_i))}\ar[r]
           &H^2(\O_X(\mathbf{Q}))\\
  H^1(\O_X)^{\oplus r}\ar[r]&\displaystyle{\bigoplus_i^{\phantom{N}}
  H^1(\O_X(\mathbf{q}_i))}\ar[r]
           &H^1(\O_X(\mathbf{Q}))\\
  H^0(\O_X)^{\oplus r}\ar[r]&\displaystyle{\bigoplus_i^{\phantom{N}}
  H^0(\O_X(\mathbf{q}_i))}\ar[r]
           &H^0(\O_X(\mathbf{Q}))
} 
\save="x"!LD+<-3mm,0pt>;"x"!RD+<0pt,0pt>**\dir{-}?>*\dir{>}\restore
\save="x"!LD+<35mm,-3mm>;"x"!LU+<35mm,-2mm>**\dir{-}?>*\dir{>}\restore
\save!CD+<0mm,-4mm>*{p}\restore
\save!UL+<32mm,-5mm>*{q}\restore
\save="x"!LD+<0mm,60mm>;"x"!LD+<120mm,0mm>**\dir{.}\restore
\end{xy} \label{eq:Tss}
\end{equation}

Fortunately it is straight-forward to compute the cohomology groups
$H^n(V,\O_X(\mathbf{v})) = H^n(X,\O_X(\mathbf{v}))$. First, the
cohomology groups $H^n(V,\O_V(\mathbf{v}))$ can be computed. Actually
we need to study these cohomology groups in detail and we give a
relevant method (if not the most efficient) in the appendix. Then one
may use exact sequences of the form
\begin{equation}
\xymatrix@1{
0\ar[r]&\O_V(\mathbf{v}-\mathbf{Q})\ar[r]&\O_V(\mathbf{v})
\ar[r]&\O_X(\mathbf{v})\ar[r]&0,
} \label{eq:Ocok}
\end{equation}
to restrict to $X$.\footnote{Perhaps the easiest way to compute
  this is to express it in terms of a module as a
  cokernel as in (\ref{eq:Ocok}) and then compute the cohomology using  
the  Macaulay 2 package 
\cite{Mac2:toric}.}
  
The dotted line in the spectral sequence (\ref{eq:Tss}) represented
terms which contribute to $H^1(T_X)$, that is, deformations of complex
structure. Since $X$ is a \CY\ threefold we know that
$H^1(\O_X)=H^2(\O_X)=0$. Also, since $\mathbf{Q}$ corresponds to an
ample divisor, $H^1(\O_X(\mathbf{Q}))=0$. The contribution to
$H^1(T_X)$ from the zeroth row (i.e., $q=0$) corresponds to the
cokernel of the $d_1$ map induced from the complex (\ref{eq:TX}). This
is given by elements of $H^0(\O_X(\mathbf{Q}))$ which are not
multiples of $\partial W/\partial x_i$. This is immediately
recognizable as deformations of the defining polynomial $W$ modulo
reparametrizations. These are the usual ``polynomial deformations'' of
$X$. The spectral sequence then yields the following little result:
\begin{theorem}
  The non-polynomial deformations of complex structure for a \CY\
  hypersurface $X$ in a toric variety are given by
\begin{equation}
 \displaystyle{\bigoplus_i^{\phantom{N}}
  H^1(\O_X(\mathbf{q}_i))}.
\end{equation} \label{th:nonpoly}
\end{theorem}
This should be compared with a similar result obtained for deformations
of complex structure of smooth projective toric varieties~\cite{Ilten:2009tv}. 
 
\subsection{$\End(T)$}  \label{ss:EndT}

We would now like to compute the cohomology groups
$H^k(X,\End(T))$. These groups may also be written $\Ext_X^1(T,T)$. The
machinery of the derived category is well-suited to compute these
cohomology groups as we discuss. We refer to \cite{GM:Hom,Wei:hom} for
more details.

Given a complex
\begin{equation}
  \cE^\bullet =
\xymatrix@1{
  \cdots\ar[r]&\cE^{-1}\ar[r]&\cE^0\ar[r]&\cE^1\ar[r]&\cdots,
}
\end{equation}
of coherent sheaves and a similar complex $\cF^\bullet$ we may form an
object in $\DC(X)$ which represents the object $\sHom(\cE,\cF)$. It is given by
the complex whose $n$th term is
\begin{equation}
\sHom(\cE^\bullet,\cF^\bullet)^n = \bigoplus_{j=i+n}\sHom(\cE^i,\cF^j).
\end{equation}
If $\phi\in\sHom(\cE^\bullet,\cF^\bullet)^n$ then we define the
differential of this new complex by
\begin{equation}
  d^n(\phi) = d_{\cF}\phi - (-1)^n\phi d_{\cE}, \label{eq:dphi}
\end{equation}
as in \cite{ST:braid}.
The total cohomology of the object $\sHom(\cE,\cF)$ represents the
``hyperext'' groups $\Ext(\cE,\cF)$.

Let us apply the above to the case of the tangent sheaf. From
(\ref{eq:TX}) we have a complex representing $\sHom(T,T)$ given by
\begin{equation}
\xymatrix@1{
\O_X(-\mathbf{Q})^{\oplus r}\ar[r]&
{\begin{matrix}\bigoplus_i\O_X(\mathbf{q}_i-\mathbf{Q})\\
  \oplus\\
  \bigoplus_i\O_X(-\mathbf{q}_i)^{\oplus r}\end{matrix}}\ar[r]&
\poso{{\begin{matrix} \O_X^{\oplus r^2}\\
\oplus\\
\bigoplus_{i,j}\O_X(\mathbf{q}_i-\mathbf{q}_j)\\
\oplus\\
\O_X\end{matrix}}}\ar[r]&
{\begin{matrix}\bigoplus_i\O_X(\mathbf{q}_i)^{\oplus r}\\
  \oplus\\
  \bigoplus_i\O_X(\mathbf{Q}-\mathbf{q}_i)\end{matrix}}\ar[r]&
\O_X(\mathbf{Q})^{\oplus r}
} \label{eq:HomTT}
\end{equation}
where the dotted line represents position zero. 
The maps in this complex are derived from $x_i\mathbf{q}_i$ and
$\partial_i W$ by using (\ref{eq:dphi}).
This yields
\begin{theorem}
  There is a spectral sequence whose $E_1$ term is given by
\begin{equation}
  E_1^{p,q} = H^q(X,\sHom(T,T)^p),
\end{equation}
where $\sHom(T,T)$ is given by (\ref{eq:HomTT}). This converges to
\begin{equation}
  H^n(X,\End(T)) \cong \bigoplus_{p+q=n} E_\infty^{p,q}.
\end{equation}  \label{th:m}
\end{theorem}

This theorem gives a practical method of computing $H^1(\End(T))$ as
we discuss in several examples below.

\subsection{The Quintic}  \label{ss:quint}

The quintic threefold in $\P^4$ provides a simple example to
demonstrate theorem \ref{th:m}. As we will see in this paper, the
quintic is deceptively simple and fails to demonstrate most of the
interesting phenomena that can happen for counting singlets. Nevertheless
it always provides a good example to start with.

For the quintic, the tangent complex (\ref{eq:TX}) becomes
\begin{equation}
\xymatrix@1{
\O_X\ar[r]^-{x_i}&\O_X(1)^{\oplus 5}\ar[r]^-{\partial_iW}&\O_X(5).
}
\end{equation}
Before writing down the full spectral sequence we should note that
Serre duality gives
\begin{equation}
  H^k(X,\O_X(\mathbf{v}) \cong H^{3-k}(X,\O_X(\mathbf{-v})).
\end{equation}
This makes the second and third row of the spectral sequence copies of
row one and zero written in reverse. 
We obtain
\begin{equation}
\begin{xy}
\xymatrix@C=11mm@R=5mm{
  &\vdots&\vdots&\vdots&\vdots&\\
\scriptstyle H^1(\O_X(-5))\ar[r]&
{\begin{matrix}\scriptstyle H^1(\O_X(-1)^{\oplus 5})\\
  \scriptstyle \oplus\\
  \scriptstyle H^1(\O_X(-4)^{\oplus 5})\end{matrix}}\ar[r]&
  {\begin{matrix} \scriptstyle H^1(\O_X)\\
\oplus\\
\scriptstyle H^1(\O_X^{\oplus25})\\
\oplus\\
\scriptstyle H^1(\O_X)\end{matrix}}\ar[r]&
{\begin{matrix}\scriptstyle H^1(\O_X(1)^{\oplus 5})\\
  \oplus\\
  \scriptstyle H^1(\O_X(4)^{\oplus 5})\end{matrix}}\ar[r]&\scriptstyle  H^1(\O_X(5))\\
\scriptstyle H^0(\O_X(-5))\ar[r]&
{\begin{matrix}\scriptstyle H^0(\O_X(-1)^{\oplus 5})\\
  \oplus\\
  \scriptstyle H^0(\O_X(-4)^{\oplus 5})\end{matrix}}\ar[r]&
  {\begin{matrix} \scriptstyle H^0(\O_X)\\
\oplus\\
\scriptstyle H^0(\O_X^{\oplus25})\\
\oplus\\
\scriptstyle H^0(\O_X)\end{matrix}}\ar[r]&
{\begin{matrix}\scriptstyle H^0(\O_X(1)^{\oplus 5})\\
  \oplus\\
  \scriptstyle H^0(\O_X(4)^{\oplus 5})\end{matrix}}\ar[r]& \scriptstyle H^0(\O_X(5))
}
\save="x"!LD+<-3mm,0pt>;"x"!RD+<0pt,0pt>**\dir{-}?>*\dir{>}\restore
\save="x"!LD+<62mm,-3mm>;"x"!LU+<62mm,-2mm>**\dir{-}?>*\dir{>}\restore
\save!CD+<0mm,-4mm>*{p}\restore
\save!LU+<59mm,-4mm>*{q}\restore
\end{xy} \label{eq:QEndT}
\end{equation}
That is,
\begin{equation}
\begin{xy}
\xymatrix@C=11mm{
  \smc{125}\ar[r]^{d_1^{(1)\vee}}&\smc{25+350}\ar[r]^{d_1^{(0)\vee}}&\smc{1+25+1}\\
  &\scriptstyle0&\scriptstyle0&\scriptstyle0\\
  &\scriptstyle0&\scriptstyle0&\scriptstyle0\\
  &&\smc{1+25+1}\ar[r]^{d_1^{(0)}}&\smc{25+350}\ar[r]^{d_1^{(1)}}&\smc{125}
} 
\save="x"!LD+<-3mm,0pt>;"x"!RD+<0pt,0pt>**\dir{-}?>*\dir{>}\restore
\save="x"!LD+<40mm,-3mm>;"x"!LU+<40mm,10mm>**\dir{-}?>*\dir{>}\restore
\save!CD+<20mm,-4mm>*{p}\restore
\save!UL+<36mm,10mm>*{q}\restore
\save="x"!LD+<0mm,41mm>;"x"!LD+<78mm,0mm>**\dir{.}\restore
\end{xy} \label{eq:QEndT1}
\end{equation}
where we show contributions to $H^1(\End(T))$ with the dotted line.

To compute the $E_2$ stage of the spectral sequence requires an
explicit determination of the $d_1$ maps in (\ref{eq:QEndT1}). This is
not so bad since the monomials of degree $n$ form a basis (up to the
quintic defining equation) of $H^0(\O(n))$.
Actually, since the holonomy of the quintic threefold is precisely $\SU(3)$, the
tangent sheaf is irreducible and so, by Schur's lemma, $H^0(\End(T))$
has dimension one. This means the map $d_1^{(0)}$ in (\ref{eq:QEndT1}) has a
one-dimensional kernel. 

The map $d _1^{(1)}$ is of the form
\begin{equation}
\xymatrix@1@C=20mm{
{\begin{matrix}H^0(\O_X(1))^{\oplus5}\\\oplus\\
H^0(\O_X(4))^{\oplus5}\end{matrix}}
\ar[r]^-{\left(\begin{smallmatrix}\partial_iW&x_i\end{smallmatrix}\right)}&
H^0(\O_X(5))
}
\end{equation}
Since clearly any degree 5 polynomial can be expressed as a
sum $\sum_i x_ig_i$, for quartic $g_i$'s, the map $d _1^{(1)}$
is surjective for any $W$.

Thus the spectral sequence degenerates at $E_2$ and we obtain
\begin{equation}
\begin{split}
\dim H^1(X,\End(T)) &= 25+350-125-1-25\\
   &= 224,
\end{split}
\end{equation}
in agreement with known results \cite{W:issues} and section
\ref{ss:lgquinitc}.  Moreover, all $224$ singlets correspond to (0,2) GLSM
deformations.


\section{Relating the Computations}  \label{s:compare}

The main point of this work lies in comparing calculations of the
singlet spectrum valid at various loci in the moduli space of (2,2)
theories.  At different loci we apply different techniques, and in
comparing the results we can find interesting relations between the
calculations.  

\subsection{$\End(T)$ and GLSM Deformations} \label{ss:endglsm}
The spectral sequence of the  section \ref{ss:EndT} is closely related to
the (0,2) GLSM holomorphic parameters studied 
in~\cite{Kreuzer:2010ph}.  In addition to the toric K\"ahler
parameters, the (0,2) superpotential is encoded by the maps in the complex
\begin{equation}
\xymatrix@1{
  0\ar[r]& \O_V^{\oplus r}\ar[r]^-{\left(\begin{matrix} \mathbf{E}_0 \\ \mathbf{E}_i
  \end{matrix}\right) }&\displaystyle{\begin{matrix} \O_V\\ \oplus \\ \bigoplus_i
  \O_V(\mathbf{q}_i) \end{matrix}} \ar[r]^-{(W, J_i)}& \O_V(\mathbf{Q})\ar[r]&0,
} \label{eq:EJ}
\end{equation}
where  $W \in H^0(\O_V(\mathbf{Q}))$ specifies the hypersurface.
In order for this to be a complex, this data must satisfy
\begin{equation}
\label{eq:Sconst}
\sum_i \mathbf{E}_i J_i +\mathbf{E}_0 W = 0.
\end{equation}
This is the famous (0,2) supersymmetry constraint.

A superpotential specified by $\mathbf{E}_i$, $J_i$ leads to the
same IR physics as one specified by $\mathbf{E}'_i$, $J'_i$ 
whenever the two are related by holomorphic field re-definitions.
These re-definitions act on various (0,2) multiplets and can
be identified with the following sections:
\begin{align}
\text{neutral chiral fields}   &:   H^0(\O_V^{\oplus r^2}), \nonumber\\
\text{charged chiral fields} &:   \oplus_i H^0(\O_V(\mathbf{q}_i)),
\nonumber\\
\text{charged Fermi fields} &: 
\oplus_{i,j} H^0(\O_V(\mathbf{q}_i - \mathbf{q}_j)).
\end{align}
These must of course be taken modulo gauge transformations and 
$\GU(1)_L$ invariance.
The data in $\mathbf{E}$, $J$ and $W$
encodes both bundle and polynomial \CY\ deformations.  As is familiar
from the monomial-divisor mirror map, the latter are nicely described 
by toric geometry~\cite{AGM:mdmm} of $V$ and the Newton polytope
for $W$.  In particular, a choice of $W$ fixes the polynomial complex
structure moduli of the \CY, as well as  re-definitions of the charged matter fields modulo gauge invariance.

Supposing we have fixed the complex structure on $X$, we
can ask about the remaining (0,2) deformations and field re-definitions.
The first order deformations $\delta \mathbf{E}^i$ and $\delta J_i$ fit
into the first position (recall that the zeroth position is marked by
the dotted line) of the complex
\begin{equation}
\xymatrix@1{
0\ar[r]&
\poso{{\begin{matrix} H^0(\O_V^{\oplus r^2})\\
\oplus\\
\bigoplus_{i,j}H^0(\O_V(\mathbf{q}_i-\mathbf{q}_j))\\
\oplus\\
H^0(\O_V)\end{matrix}}}\ar[r]&
{\begin{matrix}\bigoplus_iH^0(\O_V(\mathbf{q}_i)^{\oplus r})\\
  \oplus\\
  \bigoplus_iH^0(\O_V(\mathbf{Q}-\mathbf{q}_i))\end{matrix}}\ar[r]&
H^0(\O_V(\mathbf{Q})^{\oplus r}) .
} \label{eq:GLSMEJ}
\end{equation}
They must satisfy $\delta\mathbf{E}^i J_i + \mathbf{E}^i \delta J_i = 0$.

This complex is just what we get for $H^0(\sHom(T,T))$, where
$\sHom(T,T)$ is given in~(\ref{eq:HomTT}), but with two important
differences.  Firstly the sheaves relevant to the GLSM are defined
over $V$, while those relevant to the geometric analysis are defined
over $X$.  This can lead to differences in the counting.  For
instance, in general $H^0(\O_V(\mathbf{q}_i-\mathbf{q}_j)) \neq
H^0(\O_X(\mathbf{q}_i-\mathbf{q}_j))$, and the latter can have
additional holomorphic sections.  When this holds, the GLSM
superpotential modulo holomorphic field re-definitions
over-parametrizes the bundle deformations, since there are
automorphisms of the NLSM that cannot be lifted to holomorphic
re-definitions of the UV theory.  We will see an example of this
phenomenon in the septic. Secondly, and more obviously, we are missing
the first two terms of (\ref{eq:HomTT}). These will vanish on $V$ but
there are examples (not in this paper) where the restriction to $X$
can give nonzero entries.

We should also note another subtlety in the comparison: there is a 
difference between counting first order solutions to the supersymmetry 
constraint (\ref{eq:Sconst}) and
demanding that it is satisfied to all orders.  The latter leads to GLSM
deformations, while the former corresponds to massless states accessible
via the GLSM.

\subsection{Mapping Geometry to the \LG\ Theory}  \label{ss:cascade}
It is interesting to compare the geometric computation of the number
of singlets with the \LG\ description. At a crude level we know that
the numbers agree, but can we make a more precise map? We can make some way
via a series of exact sequences as follows. {\em We restrict attention
  to the case $r=1$, i.e., there is only one $\GU(1)$ charge.}

Define $\cA$ by
\begin{equation}
\xymatrix@1{
0\ar[r]&\cA\ar[r]&\bigoplus_i\O_X(\mathbf{q}_i)\ar[r]^-{\partial_i
  W}&\O_X(\mathbf{Q})\ar[r]&0.
}
\end{equation}
(\ref{eq:TX}) gives an exact sequence
\begin{equation}
\xymatrix@1{
0\ar[r]&\O_X\ar[r]&\cA\ar[r]&T\ar[r]&0.
}
\end{equation}
From these sequences, the following is exact:
\begin{equation}
\xymatrix{
0\ar[r]&\Ext^1(T,\cA)\ar[r]&\Ext^1(\cA,\cA)\ar[r]&
H^1(T)\ar[r]&H^0(\cA)\ar@{=}[d]\ar[r]&0,\\
&&&&\C
}
\end{equation}
as is
\begin{equation}
\xymatrix@1{
0\ar[r]&H^2(T)\ar[r]&\Ext^1(T,\cA)\ar[r]&\Ext^1(T,T)\ar[r]&0.
}
\end{equation}
These exact sequences show how the three sources of singlets,
$H^1(T)$, $H^2(T)$ and $\Ext^1(T,T)$ combine into $\Ext^1(\cA,\cA)$
(minus one).

Now let $L_1$ and $L_3$ be the ``universal'' contributions from the
\LG\ theory at $k=1$ and $k=3$ respectively from table \ref{t:LGuni}. 
That is
\begin{equation}
\xymatrix@1{
0\ar[r]&H^0(\O)\ar[r]&{\begin{matrix}\bigoplus_iH^0(\O(\mathbf{q}_i))\\\oplus\\
\bigoplus_{i,j}H^0(\O(\mathbf{q}_i-\mathbf{q}_j))\end{matrix}}
\ar[r]&\bigoplus_iH^0(\O(\mathbf{Q}-\mathbf{q}_i))
\ar[r]&L_1\ar[r]&0
}
\end{equation}
is exact, and $L_3=\bigoplus_iH^0(\O(\mathbf{q}_i))$. These statements
are written for the toric variety $V$ but one can show, in the case
$r=1$, that they are also valid on restriction to $X$.

The relationship between $L_1$, $L_3$ and
$\Ext^1(\cA,\cA)$ is expressed by the fact that the following two
sequences are exact (for generic $W$):
\begin{equation}
\xymatrix{
0\ar[r]&\C\ar[r]\ar@{=}[d]&
{\begin{matrix}L_3\\\oplus\\
\bigoplus_{i,j}H^0(\O_X(\mathbf{q}_i-\mathbf{q}_j))\end{matrix}}
\ar[r]&\bigoplus_iH^0(\O_X(\mathbf{Q}-\mathbf{q}_i))\ar@{=}[d]\ar[r]&
L_1\ar[r]&0\\
0\ar[r]&\C\ar[r]&
{\begin{matrix}H^0(\O_X)\\\oplus\\
\bigoplus_{i,j}H^0(\O_X(\mathbf{q}_i-\mathbf{q}_j))\end{matrix}}
\ar[r]&\bigoplus_iH^0(\O_X(\mathbf{Q}-\mathbf{q}_i))\ar[r]&
\Ext^1(\cA,\cA)\ar[r]&0
}
\end{equation}
It is easy to see from this that dimensions are correct, i.e.,
the singlet count between the \LG\ picture and the large radius
picture agree: 
\begin{equation}
  \dim L_1+\dim L_3 = \dim H^1(T) + \dim H^2(T) + \dim H^1(\End(T)),
\end{equation}
but we also see that the precise mapping of singlets between these
pictures is quite subtle.

\subsection{Orbifolds and a (0,2) McKay Correspondence}\label{ss:McK} 

Phases in which $X$ acquires orbifold singularities are an interesting
intermediate situation between the large-radius geometric phase and
the \LG\ phase.   Near an orbifold limit point we can distinguish
untwisted ``bulk'' states from twisted states localized near the
singular locus.  Deep in the orbifold phase, when all sizes in $X$
other than the cycle whose shrinking is responsible for the
singularity are taken large, the geometry near the singular locus
tends to a limit in which the space transverse to the singular locus
is simply $\C^{d-D}/\Gamma$ for a quotient group $\Gamma$ and a
singular locus of dimension $D$.  In the orbifold limit, the theory
acquires a discrete quantum symmetry, and states can be
classified by their transformation properties under it.  Invariant
states, also termed untwisted states, correspond to strings occupying
the ``bulk'' of $X$.  Charged, or ``twisted'' states represent strings
localized near the singular locus.  The spectrum of massless twisted
states can be determined from the local structure of $X$ near the
singular locus. 

When $X$ is a hypersurface in a toric variety, the GLSM provides a
simple description of the untwisted sector.  Deep in the orbifold
phase, some of the chiral fields acquire large expectation values,
breaking the gauge group down to a subgroup $\GU(1)^{r'}\times\Gamma$,
$r'<r$.
Fluctuations of these fields acquire large masses through the Higgs
mechanism\footnote{As usual, we gloss over the subtleties of this in
  two dimensions.}, and integrating them out we find an effective
theory of the remaining chiral fields interacting with $r'$ gauge
multiplets and via an effective superpotential $\hat W$.  Applying the 
GLSM picture of section~\ref{ss:endglsm} to this reduced model
produces singlets in the untwisted sector of the model.

To describe the twisted sector we use the fact that twisted states are
localized near the singular locus.  This allows us to use the local
geometry to find a free-field description near a point on the singular
locus following \cite{DHVW:}.  For the reader's convenience we recall
the analysis, recast in the notation we use here, restricting
attention for simplicity to the case $\Gamma = \Z_p$.  Near a point on
the singular locus we pick local coordinates $x_i,\ i=1\ldots 3$, on
which the $\Gamma$ action is generated by $x_i\mapsto e^{2\pi i n_i/p}
x_i$ with $\sum n_i = 0 \pmod p$, and consider a free theory with
chiral supermultiplets $\Phi_i$ and this (non-$R$) action.  The
twisted sectors of this orbifold of a free field theory, our
approximation to the twisted sectors of the orbifold phase of $X$, can
be described using the techniques of Section~\ref{s:LG} simply setting
$W=0$ and $\alpha_i=0$.   We find twisted
sectors labeled by $(k,s)$ where $k=1,\ldots p-1$, and $s=0,1$
distinguishes R ($s=0$) sectors from NS ($s=1$) boundary conditions on
the left-moving fermions.  In the
sector $(k,s)$ the boundary conditions on the fields are given by
\begin{align}
\label{eq:freenus}
\nu_i &= \frac{kn_i}{p}\pmod 1 &0<\nu_i\le 1,\nonumber\\
\nut_i &= \frac{kn_i}{p}-\frac{s}{2}\pmod 1 &-1<\nut_i\le 0,
\end{align}
while $h_i=0$ and $\htld_i=\ff{1}{2}$.
With $W=0$ computing $\Qb$ cohomology is trivial,
but we need to perform the projection onto $\Gamma$-invariant states, as well as the GSO projection.

The quotient preserves the four worldsheet supercharges given to
within overall factors by
\begin{equation}
G^- = \sum_i\gamma_i\p\bar x_i,\quad G^+ = \sum_i\gammab_i\partial x_i,\quad
\bar G^- = \sum_i\psi_i\bar \p \bar x_i,\quad\bar G^+ =
\sum_i\bar\psi_i\bar \p x_i
\end{equation}
The superscripts label the charges under the fermion number currents
given by 
\begin{equation}
\begin{split}
J &= \sum_i J_i,\quad J_i= \gammab_i\gamma_i,\\
\bar J &= \sum_i \bar J_i,\quad \bar J_i= \bar\psi_i\psi_i
\end{split}
\end{equation}

The case $D=0$---an isolated singular point---is simplest.  We study
an example in section~\ref{ss:septic}.  The free field theory
exhibits (in general) an unbroken $\GE_6$ gauge symmetry, and the
methods we presented lead to predictions for the massless spectrum in
twisted sectors.  We find $\rep{27}$ and $\brep{27}$ multiplets and
their conjugates along with the $\GE_6$ singlets related to them by
the left-moving supersymmetry which are $(2,2)$ moduli.
These are the subject of the McKay correspondence.
In our conventions, chiral $\rep{27}$s correspond to K\"ahler
deformations, while chiral $\brep{27}$s correspond to deformations of
complex structure.  There will also be $\GE_6$ singlets in the twisted
sectors, for which one can attempt to find an analogous
correspondence.  Note that the conjugate states will arise in the
conjugate sector, so that in general some of these singlet states might
be lifted in pairs by deformations resolving the singularity.

In the case $D=1$---a curve $C$ of singularities---we proceed in stages.  Near a
point on $C$ we can find local coordinates as above, such that
$n_1=0$.  There are bosonic
$x_1$ zero modes in the twisted sectors, so our states will locally be
described by functions of $x_1$.   A novel feature of this
construction will be that excluding $x_1$ means the zero mode of
$\bar\psi_1$ is no longer $\bar Q$-exact, and we will need to include
it in our computations (recall that we compute in the right-moving
R sector where $\psi_1$ satisfies the same untwisted boundary
conditions as $x_1$).
The free-field quotient preserves an unbroken $\GE_7$ gauge
symmetry given by the embedding 
$\GE_7\times\SU(2)\subset\GE_8$.\footnote{Pedants will note the
  missing quotient by $\Z_2$ here.}

We can then consider this quotient structure fibered over the 
(assumed to be) large
curve $C$.  The low-energy physics will be described by a nonlinear
sigma model on $C$ with the fields determined from the orbifold
construction as above.  These will couple to the spin connection on
$C$ so that the functions of $x_1$ become sections of appropriate
bundles.  To find the spin we note that the curvature of $C$ breaks
$\GE_7\to\GE_6$. Since this fits into the maximal embedding
$\GE_6\times\SU(3)\subset\GE_8$ we see that $\GU(1)_C$ must be an
$\SU(3)$ subgroup.  The fact that it must also commute with an $\SU(2)$
subgroup acting in the transverse directions (before taking the
quotient) determines
\begin{equation}
J_C = -J_1 - \bar J_1 + \ff12\left(J_2+J_3+\bar J_2+\bar J_3\right) \ .
\end{equation}

Simply compactifying the quotient theory on the curve $C$ will not
lead to a supersymmetric spectrum.  In fact, the local structure is
not described as $C\times\left(\C^2/\Gamma\right)$ but as a fibration
over $C$ such that the resulting space is \CY .  In terms of our
fields, the transverse coordinates $x_i$ for $i\ne 1$ transform under
$J_C$ in such a way that two of the supercharges are invariant.
Moreover, the zero mode of $\bar\psi_1$, lying in the cokernel of
$\bar Q$ in the nonlinear sigma model, becomes a section of $\Omega$,
the cotangent bundle of $C$.  When acting on a vertex operator
corresponding to a space-time fermion in a massless chiral multiplet
(with $\bar q=-\ff{1}{2}$) $\bar\psi_1$ will create the vertex operator for
a massless antichiral multiplet ($\bar q=\ff{1}{2}$).  Thus fields in the
nonlinear theory with charge $\bar q=-\ff{1}{2}$ and spin $q_C$ will produce
massless chiral multiplets corresponding to $H^0(C,\Omega^{\otimes
  q_C})$.  Fields with charge $\bar q=\ff{1}{2}$ and spin $q_C$ will produce
massless antichiral multiplets corresponding to
$H^1(C,\Omega^{\otimes(q_C+1)})$.  CPT invariance, the requirement
that each chiral multiplet of charge $q_C$ be accompanied by an
antichiral multiplet of charge $-q_C$ is then tantamount to Serre
duality on $C$.

This gives a matching between counting singlets in the orbifold
language and the large radius language which works almost perfectly.
Some interesting subtleties concerning dependence on complex structure will
be seen in section \ref{sss:oct-orb}.

\begin{figure}
\hspace{10mm}
\begin{picture}(100,50)
\put(10,50){\framebox{Large Radius Limit}}
\put(95,50){\framebox{Orbifold}}
\put(80,10){\framebox{\LG\ Theory}}
\put(48,51){\vector(1,0){46}}
\put(94,51){\vector(-1,0){46}}
\put(104,48){\vector(0,-1){33}}
\put(104,15){\vector(0,1){33}}
\put(60,53){(0,2) McKay}
\put(105,30){Cascade}
\put(55,25){\vbox{\em K\"ahler form\\ moduli space}}
\end{picture}
\caption{Mapping between \CY\ , \LG\  and orbifold phases.} \label{f:map}
\end{figure}

The correspondence between the \LG\ picture and the large radius
picture proceeds typically as shown in figure \ref{f:map}. The
orbifold corresponds to a weighted projective space where we
essentially just consider a single $\C^*$-action (i.e., $r'=1$), and so
the results of section \ref{ss:cascade} apply. That is, we may relate
the \LG\ picture to the orbifold picture using the $L_1$ and $L_3$
contributions to the cascade. Then the orbifold may be related to the
large radius limit by using the above (0,2)-McKay correspondence. 


\section{Examples}  \label{s:eg}

\subsection{The sextic in $\P^4_{\{2,1,1,1,1\}}$}

\subsubsection{Geometry}

The weighted projective space $\P^4_{\{2,1,1,1,1\}}$ with homogeneous
coordinates $[x_0,\ldots,x_4]$ has a terminal singularity at
$x_1=x_2=x_3=x_4=0$. If $X$ is a generic hypersurface of degree 6 then
it will not intersect this singularity. 
The Hodge numbers of $X$ are $h^{1,1}(X)=1$ and $h^{2,1}(X)=103$.
The analysis of
$H^1(X,\End(T))$ looks quite similar to that of the quintic. The
spectral sequence is
\begin{equation}
\begin{matrix}\vspace{20mm}\\E_1^{p,q}:\end{matrix}
\begin{xy}
\xymatrix@C=5mm@R=5mm{
  \smc{129}\ar[r]&\smc{393}\ar[r]&\smc{35}\\
  &\scriptstyle0&\scriptstyle0&\scriptstyle0\\
  &\scriptstyle0&\scriptstyle0&\scriptstyle0\\
  &&\smc{35}\ar[r]&\smc{393}\ar[r]&\smc{129}
} 
\save="x"!LD+<-3mm,0pt>;"x"!RD+<0pt,0pt>**\dir{-}?>*\dir{>}\restore
\save="x"!LD+<23mm,-3mm>;"x"!LU+<23mm,2mm>**\dir{-}?>*\dir{>}\restore
\save!CD+<20mm,-4mm>*{p}\restore
\save!UL+<21mm,3mm>*{q}\restore
\end{xy}
\end{equation}
This gives $\dim H^1(X,\End(T))=230$.  As in the case of the quintic, each
of these corresponds to a (0,2) GLSM deformation.  Adding in singlets from
$h^{1,1}$, $h^{2,1}$ and $\GU(1)$ partners we would predict a value
for the Gepner model of $230+1+103+4=338$. The actual value as given
in the table of \cite{Lutken:1988hc} is 344. We are short by 6.

\subsubsection{The \LG\ locus}
The superpotential is a degree $6$ polynomial with weights $\alpha_0 = \ff{1}{3}$
and all other $\alpha_i = \ff{1}{6}$ (we take $i=1,\ldots,4$).   Working
sector by sector, we find the following zero energy states with
$q=0$. 
 \begin{equation}
\xymatrix@C=40mm@R=5mm{
  \bar q=-\ff32&\bar q=-\ff12\\
{\begin{matrix}
\gamma_0\gammab_0 |1\ra_1\oplus\gamma_i \gammab_j |1\ra_{16}\\\oplus\\
F^i_{[1]}\gamma_i\gammab_0|1\ra_{16}\\\oplus\\
G_{[1]}^i \rho_i |1\ra_{16}\oplus H_{[2]} \rho_0|1\ra_{11}
\end{matrix}}
\ar[r]^-\Qb&
{\begin{matrix}
F_{[4]} \gamma_0 |1\ra_{46}\\\oplus\\
F_{[5]}^i \gamma_i|1\ra_{320}\end{matrix}
}\\
&x_i \gammab_j |3\ra_{16}\oplus F_{[2]} \gammab_0 |3\ra_{11}
\\
&\rho_0 \gammab_i \gammab_j |5\ra_6
}
\end{equation}

In the $k=1$ sector $\Qb : U_{-3/2} \to U_{-1/2}$ has a
one-dimensional kernel for generic $W$; the dimension increases to $5$
at the Fermat point.  This is just what we have already observed in
the quintic.  So, we find $307$ $k=1$ chiral singlets for generic $W$
and $4$ more at the Fermat point.

The $k=3$ states arise from the ``cascade'' picture described above,
while the six $k=5$ states account for the discrepancy with geometry.  As we
will see, mirror symmetry shows that these states are lifted once we turn
on the K\"ahler modulus to move away from the \LG\ point.

\subsubsection{Mirror Symmetry and K\"ahler Deformations}

The Gepner model in this moduli space is given by a (quotient of) the
product  $A_2\oplus A_5^{\oplus 4}$ of minimal models at level $k_i=4$ and 
$k_0=1$.  In addition to the universal
``cascade'' states in the $k=1$ and $k=3$ sectors, there are six
singlet states in the $k=5$ sector.  The Gepner model enjoys a
discrete $S_6$ symmetry which permutes these singlet states. 
The K\"ahler deformation does not break this symmetry, so all
of these singlets will be lifted by the deformation or none will.

Explicitly, the six massless singlet states at $k=5$ are given by
permutations of
\begin{equation}
  S = \lqs{\pz0}{-4}2 \lqs{\pz2}{-2}0 \lqs{\pz2}{-2}0 
           \lqs{\pz1}{-3}2 \lqs{\pz1}{-3}2\Big|
  \lqs 011 \lqs 231 \lqs 231 \lqs 121\lqs 121 
\end{equation}
For clarity we have not cast this into standard form, exhibiting it in
a way that makes the twist manifest.

The mirror model is given by a $\Z_6^2\times\Z_3$ quotient represented
by the vectors
\begin{equation}
m^{(1)} = (0,1,0,0,-1)\quad m^{(2)} = (0,0,1,0,-1)\quad  m^{(3)} = (1,0,0,0,-2)\ .
\end{equation}
The state mirror to $S$
(obtained by reversing the signs of $q$ and $s$) satisfies 
\begin{align}
\bar q = q + m^{(1)} + m^{(2)} - m^{(3)} - m^{(0)},
\end{align}
where $m^{(0)}=(1,1,1,1,1)$, and
so will appear in the sector $(k;t) = (11;1,1,2)$.  Note that while
the mirror model obviously shares the permutation symmetry of the
original, this is not evident in its presentation as a quotient.  Thus
the states related to $S$ by permutations will arise in other twisted
sectors. 

In the \LG\ model the discrete group acts via
\begin{equation}
 w^1 = (0,\ff{1}{6},0,0,-\ff{1}{6})\quad w^2 = (0,0,\ff{1}{6},0,-\ff{1}{6})\quad w^3=
 (\ff{1}{3},0,0,0,-\ff{1}{3})\ . 
\end{equation}
Constructing the \LG\ orbifold we find in this sector the following
states at $q=0$:
\begin{equation}
\xymatrix@C=40mm@R=5mm{
  \bar q=-\ff12&\bar q=\ff12\\
\gammab_0 x_1^2 x_2^2\rho_3^2\rho_4^2|v\ra
\ar[r]^-\Qb&
x_1^3 x_2^3\rho_3\rho_4|v\ra
}
\end{equation}
where $|v\ra$ is the twisted vacuum.  To find the action of $\Qb$ note
that since $\nut_0 =0$,  the only term that can possibly contribute 
is $\gammab_0^\dagger(\p_0 W)_0$.  At the Gepner point
$\p_0 W=2 x_0^2$, and expanding this we find that 
the $\qb=-\ff{1}{2}$ state
contributes to the cohomology at the Gepner point.

The mirror \LG\ model has a unique superpotential deformation (related
by mirror symmetry to the K\"ahler deformation of the original model).
This corresponds to adding to the Fermat superpotential the unique
monomial invariant under the quotient group:
\begin{equation}
\delta W = \psi x_0 x_1 x_2 x_3 x_4\ .
\end{equation}
This modifies $\Qb$ as found above, introducing a term 
$\gammab_0^\dagger\psi x_1 x_2 x_3 x_4$, which upon expansion gives 
\begin{equation}
\delta \Qb = \cdots + \psi\gammab_0^\dagger x_1
x_2\rho_3^\dagger\rho_4^\dagger \ ,
\end{equation}
rendering the kernel trivial.  The state $S$, and thus all
six $k=5$ singlets states found above, are lifted for 
$\psi\neq 0$ by a K\"ahler dependent mass term.

\subsection{The septic in $\P^4_{\{3,1,1,1,1\}}$}\label{ss:septic}
\subsubsection{Geometry}

Let $X_{\textrm{orb}}$ be a septic hypersurface in
$\P^4_{\{3,1,1,1,1\}}$. This weighted projective space has a
codimension 4 quotient singularity but the degree of the hypersurface
forces $X_{\textrm{orb}}$ to pass through this
point. $X_{\textrm{orb}}$ thus has an isolated singularity of the form
$\C^3/\Z_3$.

The polytope in the $N$ lattice associated with $\P^4_{\{3,1,1,1,1\}}$
is not reflexive but we may analyze the septic hypersurface in the
general setup of the gauged linear $\sigma$-model in the
following way. Let $P^\circ$ be the Newton polytope in $M$ for septics
in $\P^4_{\{3,1,1,1,1\}}$. The 8 vertices are given by the equation
\begin{equation}
  x_0^2x_1+x_0^2x_2+x_0^2x_3+x_0^2x_3+
  x_1^7+x_2^7+x_3^7+x_3^7.
\end{equation}
Then define $P\subset N$ as the polar
polytope. $P$ has 6 vertices, which may be taken to be $(1,0,0,0)$,
$(0,1,0,0)$, $(0,0,1,0)$, $(0,0,0,1)$, $(-3,-1,-1,-1)$ and
$(-1,0,0,0)$. 

One then constructs a toric
variety $V$ associated to $P$ with following data. The homogeneous
coordinate ring is $R=[x_0,\ldots,x_5]$, the charges are given by
\begin{equation}
  \Phi = \begin{pmatrix} 0&1&1&1&1&-3\\
         1&0&0&0&0&1\end{pmatrix},
\end{equation}
and let $B$ be $(x_0,x_5)(x_1,x_2,x_3,x_4)$. $V$ is then a
$\P^1$-bundle over $\P^3$. $X$ is a smooth \CY\ hypersurface in
this. The vertices of the Newton polytope correspond to the equation
\begin{equation}
  x_0^2x_1+x_0^2x_2+x_0^2x_3+x_0^2x_3+
  x_1^7x_5^2+x_2^7x_5^2+x_3^7x_5^2+x_3^7x_5^2.  \label{eq:h7}
\end{equation}
The divisor $x_5=0$ corresponds to a $\P^2\subset X$ with normal
bundle $\O(-3)$ which arises as the exceptional divisor of the
resolution of the $\C^3/\Z_3$ quotient singularity in
$X_{\textrm{orb}}$.
One easily computes $h^{1,1}(X)=2$ and $h^{2,1}(X)=122$.

$X$ has an interesting relation with the \CY\ threefold $X'$, the
resolution of the degree 14 hypersurface in $\P^4_{\{7,2,2,2,1\}}$ as
first observed in \cite{Candelas:1994bu}.  One may follow extremal
transitions between hypersurfaces in toric varieties \cite{ACJM:srch}
by shrinking the Newton polytope and thus growing its polar. That is,
one drops terms in the defining equation corresponding to vertices of
the convex hull of $P^\circ$. Typically this makes $X$ singular. The
fact that $P$ grows corresponds to a resolution of singularities which
allows $X$ to pass through an extremal transition. We may try to do
the same thing with our septic by shrinking the convex hull of
(\ref{eq:h7}) to that of
\begin{equation}
  x_0^2x_1+
  x_1^7x_5^2+x_2^7x_5^2+x_3^7x_5^2+x_3^7x_5^2. \label{eq:h7p}
\end{equation}
The polar of this Newton polytope corresponds to
$\P^4_{\{7,2,2,2,1\}}$ and thus $X$ appears to have undergone an
extremal transition to $X'$. That said, the defining equation
(\ref{eq:h7p}) is actually {\em smooth}. The supposed transition is
not a transition at all and $X'$ is merely a smooth deformation of
$X$.

Actually the septic $X$ should be considered more generic than the
degree 14 hypersurface $X'$ in the following sense. All 122
deformations of complex structure of $X$ are seen as polynomial
deformations. For $X'$, theorem \ref{th:nonpoly} shows that 15 of the
deformations are non-polynomial. This is because $X'$ has an
exceptional divisor of the form $C\times\P^1$ where $C$ is a genus 15
curve. As observed in \cite{Wil:Kc}, a generic deformation of $X'$
will break this divisor up into 28 rational curves. The latter
geometry is seen in a generic $X$.

The analysis of
$H^1(X,\End(T))$ proceeds as follows. The
spectral sequence is
\begin{equation}
\begin{matrix}\vspace{20mm}\\E_1^{p,q}:\end{matrix}
\begin{xy}
\xymatrix@C=5mm@R=5mm{
  \smc{316}\ar[r]&\smc{690}\ar[r]&\smc{87}\\
  &\scriptstyle0&\scriptstyle0&\scriptstyle0\\
  &\scriptstyle0&\scriptstyle0&\scriptstyle0\\
  &&\smc{87}\ar[r]&\smc{690}\ar[r]&\smc{316}
} 
\save="x"!LD+<-3mm,0pt>;"x"!RD+<0pt,0pt>**\dir{-}?>*\dir{>}\restore
\save="x"!LD+<23mm,-3mm>;"x"!LU+<23mm,2mm>**\dir{-}?>*\dir{>}\restore
\save!CD+<20mm,-4mm>*{p}\restore
\save!UL+<21mm,3mm>*{q}\restore
\end{xy}
\end{equation}
This has a new feature compared to the quintic and sextic of the
previous sections. The $H^0$ of various line bundles computed in the
bottom row of the spectral sequence must be computed on $X$ and not
copied from $V$. For example $H^0(\O_V(4,-1))$ is trivial while
$H^0(\O_X(4,-1))$ is dimension one.

Anyway, this gives $\dim H^1(X,\End(T))=288$.  We can compare this to
the Gepner model for $X'$.  Adding in singlets from $h^{1,1}$,
$h^{2,1}$ and $\GU(1)$ partners we would predict a value for the
Gepner model of $288+2+122+3=415$, which is correct.

The counting of (0,2) GLSM deformations following~\cite{Kreuzer:2010ph}
na\"\i vely yields $292$ parameters associated to $H^1(X,\End(T))$.  
This is due to the four extra 
automorphisms counted by $H^0(\O_X(4,-1))$ that cannot be lifted to the
GLSM.  Modulo this subtlety, we expect that all of the (0,2) singlets identified
by the geometric computation can be integrated up to deformations of the
(0,2) superpotential.

\subsubsection{The Orbifold}   \label{ss:orb7}

As argued in section \ref{ss:McK}, we may analyze $X_{\textrm{orb}}$ in terms of
the toric picture of the weighted projective space. That is, we have a
homogeneous coordinate ring $R=\C[x_0,\ldots,x_4]$ with the grading
giving by the weights $(3,1,1,1,1)$ and the irrelevant ideal is 
simply $B=(x_0,x_1,\ldots,x_4)$.

The spectral sequence of (\ref{eq:Tss}) which computes the cohomology
of the tangent sheaf becomes
\begin{equation}
\begin{matrix}\vspace{20mm}\\E_1^{p,q}:\end{matrix}
\begin{xy}
\xymatrix@C=5mm@R=5mm{
  \smc{}\\
  \scriptstyle0&\scriptstyle0\\
  \scriptstyle0&\scriptstyle0\\
  \smc{}\ar[r]&\smc{37}\ar[r]&\smc{158}
} 
\save="x"!LD+<-3mm,0pt>;"x"!RD+<0pt,0pt>**\dir{-}?>*\dir{>}\restore
\save="x"!LD+<6mm,-3mm>;"x"!LU+<6mm,2mm>**\dir{-}?>*\dir{>}\restore
\save!CD+<20mm,-4mm>*{p}\restore
\save!UL+<4mm,3mm>*{q}\restore
\end{xy}
\end{equation}
which predicts $h^{2,1}=122$ and $h^{1,1}=1$. Recall that these are
the contributions from the {\em untwisted\/} sector of the orbifold.
The value of $h^{2,1}$ is correct but we need to add one twisted state
to $h^{1,1}$ to account for the $\C^3/\Z_3$ singularity. Then we agree
with above.

The corresponding spectral sequence for $\End(T)$ gives
\begin{equation}
\begin{matrix}\vspace{20mm}\\E_1^{p,q}:\end{matrix}
\begin{xy}
\xymatrix@C=5mm@R=5mm{
  \smc{158}\ar[r]&\smc{496}\ar[r]&\smc{59}\\
  &\scriptstyle0&\scriptstyle0&\scriptstyle0\\
  &\scriptstyle0&\scriptstyle0&\scriptstyle0\\
  &&\smc{59}\ar[r]&\smc{496}\ar[r]&\smc{158}
} 
\save="x"!LD+<-3mm,0pt>;"x"!RD+<0pt,0pt>**\dir{-}?>*\dir{>}\restore
\save="x"!LD+<23mm,-3mm>;"x"!LU+<23mm,2mm>**\dir{-}?>*\dir{>}\restore
\save!CD+<20mm,-4mm>*{p}\restore
\save!UL+<21mm,3mm>*{q}\restore
\end{xy}
\end{equation}
which yields $\dim H^1(X_{\textrm{orb}},\End(T))=280$. Comparing to
above, we see that there must be 8 twisted stated to yield to the
total of 288.

So we predict that there are 9 twisted singlets states --- 1
contributing to $h^{1,1}$ and 8 to $H^1(\End(T))$.  Indeed the
free-field calculation of \cite{DHVW:} reproduces this.  We have
here two twisted sectors $k=1,2$ and since they are related by
conjugation we may restrict attention to $k=1$.  These twisted
vacua are invariant under $\Z_3$, i.e. $q_g$ of~(\ref{eq:orbcharge}) is
zero.

In the $(k,s)=(1,0)$  (twisted R) sector we find $\nu_i=\ff{1}{3},\ \nut_i=-\ff{2}{3}$. As usual
$E_{(1,0)}=0$, and the vacuum charges are $(q_{(1,0)},\bar q_{(1,0)}) = (\ff{1}{2},-\ff{1}{2})$.
There are no zero modes in the internal theory, so the unique R ground
state leads (after GSO projection) to chiral fermions in the
$\rep{16}_{1/2}$ of $\SO(10)$.

In the $(1,1)$ (twisted NS) sector we have $\nu_i=\ff{1}{3},\ \nut_i = -\ff{1}{6}$.  Here
$E_{(1,1)} = -\ff{1}{2}$ and the vacuum charges are $(q_{(1,1)},\bar q_{(1,1)}) =
(-1,-\ff{1}{2})$.  There are thus massless states in the $\rep{10}_{-1}$.
Corresponding to these is the associated singlet (K\"ahler modulus)
given by $G^+|1,1\ra$.  Internal excitations yield the nine
massless singlet states $x_i\gammab_j|1,1\ra$ with charge $(0,-\ff{1}{2})$ of
which one linear combination is the K\"ahler modulus mentioned above, as
well as the state $\gammab_1\gammab_2\gammab_3|1,1\ra$ with charges
$(2,-\ff{1}{2})$.

We thus find a chiral $\rep{27}$ predicted by the McKay
correspondence, as well as the 9 singlets predicted above.  

\subsubsection{The \LG\ analysis}\label{ss:LG7}
The \LG\ phase of the corresponding GLSM is described by a degree $7$
superpotential $W(x_0,\ldots, x_4)$, with $\alpha_0 = \ff{3}{7}$ and 
$\alpha_i = \ff{1}{7}$ for $i = 1,\ldots, 4$.  The zero energy states with $q=0$
are given by
 \begin{equation}
\xymatrix@C=30mm@R=5mm{
  \bar q=-\ff32&\bar q=-\ff12&\bar q=\ff12\\
{\begin{matrix}
\gamma_0\gammab_0|1\ra_1\oplus\gamma_i\gammab_j |1\ra_{16}\\\oplus\\
G^j_{[1]}\rho_j |1\ra_{16}\\\oplus\\
H_{[2]}^{i0} \gamma_i \gammab_0|1\ra_{40}\oplus G^0_{[3]} \rho_0 |1\ra_{21}
\end{matrix}}
\ar[r]^-\Qb&
{\begin{matrix}
H^0_{[4]} \gamma_0 |1\ra_{39}\\\oplus\\
H_{[6]}^i \gamma_i|1\ra_{420}\end{matrix}
}\\
\rho_0 \gammab_i|3\ra_4\ar[r]^-{\Qb}&
{\begin{matrix}
K^j_{[1]}\gammab_j\gamma_0|3\ra_{16}\\\oplus\\
K^0_{[3]}\gamma_0|3\ra_{21}\end{matrix}}
\\
&{\begin{matrix}x_0 \gammab_0|5\ra_1\\\oplus\\
x_i\gammab_j |5\ra_{16}\end{matrix}}\ar[r]^-{\Qb}&
x_0^2x_i |5\ra_4
}
\end{equation}

Once again, for $k=1$, $\Qb: U_{-3/2} \to U_{-1/2}$ has a
one-dimensional kernel for $W$ generic and a five-dimensional kernel
at the Gepner point, which corresponds to,
\begin{align}
W = (x_0^2x_1 + x_1^7)+x_2^7+x_3^7+x_4^7,
\end{align}
i.e. the minimal model $D_8\oplus  A_6^{\oplus 3}$.
Adding up the states, we find $366$ $k=1$ chiral singlets for generic $W$.  

In the $k=3$ sector, $\Qb$ has a trivial kernel unless $W^{55} =
W^{i5} = 0$, but that is a singular superpotential.  So, this sector
contributes $16+21-4 = 33$ singlets for any non-singular $W$.
This model demonstrates the ``cascade'' described in \ref{ss:lguni}.
In the third sector we have in addition to the $\qb=-\ff{1}{2}$ states
listed in (\ref{eq:uni3}) the set of four states of charge
$\qb=-\ff{3}{2}$ and a nontrivial action of $\Qb$ so that there are less
massless singlets in this sector than the ``universal'' prediction.
However, these states return in the $k=9$ sector (not shown) as four
states of charge $\qb=-\ff{1}{2}$ (related by conjugation to the four states
at charge $\qb=\ff{1}{2}$ at $k=5$).

Finally, we consider the $k=5$ sector. Writing $W$ as
\begin{align}
W = f_{[7]}(x_i) + g_{[4]}(x_i) x_0 + s^i x_i  x_0^2,
\end{align}
we see that the action of $\Qb$ on an arbitrary state 
\begin{equation}
|\psi\ra = (a x_0\gammab_0 + b^{ij} x_i \gammab_j ) |5\ra \in U_{-1/2} 
\end{equation}
is
\begin{equation}
\Qb |\psi\ra = (2a s^i x_i +s^i b^{ji} x_j ) x_0^2 |5\ra. 
\end{equation}
Thus, all of the $U_{1/2}$ states are $\Qb$-exact for any non-singular
$W$, and we find $13$ chiral singlets in $k=5$. This total agrees with
the orbifold and large radius phases.

\subsection{The octic in $\P^4_{\{2,2,2,1,1\}}$}\label{ss:octic}
\subsubsection{Geometry}

The weighted projective space $\P^4_{\{2,2,2,1,1\}}$ with homogeneous
coordinates $[x_0,\ldots,x_4]$ has a $\Z_2$ quotient singularity along
$x_0=x_1=x_2=0$. This may be resolved to yield a toric variety $V_0$
with homogeneous coordinates $[x_0,\ldots,x_5]$, an irrelevant ideal
$B=(x_0,x_1,x_2,x_5)(x_3,x_4)$ and grades given by the charge matrix 
\begin{equation}
  \Phi = \begin{pmatrix} 0&0&0&1&1&-2\\
         1&1&1&0&0&1\end{pmatrix}. \label{eq:wts21}
\end{equation}

$X_0$ is an octic hypersurface in $V_0$ with defining equation, in Fermat
form,
\begin{equation}
  x_0^4+x_1^4+x_2^4+(x_3^8+x_4^8)x_5^4.  \label{eq:Frm21}
\end{equation}
Mirror symmetry was studied in detail for this example in
\cite{CDFKM:I}.

The exceptional set in $X_0$ formed by the $\Z_2$-resolution is of the
form $E=C\times\P^1$, where $C$ is a genus 3 curve. One has
$h^{1,1}(X)=2$, where the two deformations of $B+iJ$ can be considered
to be the overall volume and a size of $E$. One may also show
$h^{2,1}=86$. Of these 86 deformations of complex structure, 83 are
obtained by deformations of the polynomial (\ref{eq:Frm21}). The
remaining 3 deformations of complex structure arise from
$H^1(\O_X(-2,1))=3$ in agreement with theorem \ref{th:nonpoly}. The 83
polynomial deformations of complex structure preserve
$E=C\times\P^1$. We will see below that the remaining 3 deformations
break $E$ apart into 4 disjoint $\P^1$'s in accord with \cite{Wil:Kc}.

The $E_1$ stage of the spectral sequence to compute $H^1(X_0,\End(T))$ is given by
\begin{equation}
\begin{matrix}\vspace{20mm}\\E_1^{p,q}:\end{matrix}
\begin{xy}
\xymatrix@C=5mm@R=5mm{
  \smc{208}\ar[r]&\smc{426}\ar[r]&\smc{40}\\
  &\smc{6}\ar[r]&\smc{15}\\
  &&\smc{15}\ar[r]&\smc{6}\\
  &&\smc{40}\ar[r]&\smc{426}\ar[r]^{d_1}&\smc{208}
} 
\save="x"!LD+<-3mm,0pt>;"x"!RD+<0pt,0pt>**\dir{-}?>*\dir{>}\restore
\save="x"!LD+<23mm,-3mm>;"x"!LU+<23mm,2mm>**\dir{-}?>*\dir{>}\restore
\save!RD+<0mm,-4mm>*{p}\restore
\save!CL+<21mm,0mm>*{q}\restore
\end{xy} \label{eq:ssE121}
\end{equation}

The map marked $d_1$ in (\ref{eq:ssE121}) fails to be surjective in
this case which makes for more interesting analysis compared to the
above examples. Let $R_4$ denote the vector space of degree 4
polynomials in the variables $\{x_0,x_1,x_2\}$. Then one can show that
\begin{equation}\label{eq:cokerdi}
  \coker d_1 = \frac{R_4}{(x_0,x_1,x_2)\left(\partial_0W,
 \partial_1W,\partial_2W\right)|_{x_5=0}}.
\end{equation}
This is a 6-dimensional space. In particular, for example, if $W$ is
in Fermat form then $\coker d_1$ is spanned by
$\{x_0^2x_1^2,x_0^2x_2^2,x_1^2x_2^2,x_0^2x_1x_2,x_0x_1^2x_2,x_0x_1x_2^2\}$.
The map on row one of the spectral sequence is shown to be surjective
in the appendix.

If one were to replace the map $x_i\mathbf{q}_i$ in (\ref{eq:TX}) with
a generic map of the right multi-degree then the map $d_1$ in
(\ref{eq:ssE121}) becomes surjective. This means that a deformation to
a more generic (0,2)-model kills any massless states that appear at
the (2,2)-locus due to a failure of surjectivity of $d_1$. For this
{\em generic\/} (0,2)-model the spectral sequence becomes degenerate
at the $E_2$ stage:
\begin{equation}
\begin{matrix}\vspace{20mm}\\E_2^{p,q}:\end{matrix}
\begin{xy}
\xymatrix@C=5mm@R=5mm{
  \scriptstyle0&\smc{179}&\smc{}\\
  &\scriptstyle0&\smc{9}\\
  &&\smc{9}&\scriptstyle0\\
  &&\smc{}&\smc{179}&\scriptstyle0
} 
\save="x"!LD+<-3mm,0pt>;"x"!RD+<0pt,0pt>**\dir{-}?>*\dir{>}\restore
\save="x"!LD+<22mm,-3mm>;"x"!LU+<22mm,2mm>**\dir{-}?>*\dir{>}\restore
\save!RD+<0mm,-4mm>*{p}\restore
\save!CL+<20mm,0mm>*{q}\restore
\end{xy} \label{eq:ssE122}
\end{equation}
This yields a generic value of $\dim H^1(X_0,\End(T))=188$.

However, on the (2,2)-locus, where $d_1$ fails to be surjective
$H^1(X_0,\End(T)$ may jump to a higher value. A precise analysis of this
is not too difficult, but the technical details may be a little
distracting. So this computation is left to the appendix. The result
is that for the Fermat polynomial one finds $\dim H^1(X_0,\End(T))=200$,
but that this value falls back to 188 for a generic W, even on the
(2,2)-locus. This kind of jumping in $\dim H^1(X_0,\End(T))$ as one
varies the complex structure was seen in other examples in
\cite{Berglund:1990rk}.

Three of the deformations of complex structure of $X_0$ are obstructed in
the sense that they prevent $X_0$ being embedded in the toric variety we
have considered so far. That said, it is still possible to understand
these deformations in terms of a hypersurface in a toric variety as
follows.

$V$ is the crepant resolution of the weighted projective space
$\P^4_{\{2,2,2,1,1\}}$. It can be viewed as a $\P^3$-bundle over
$\P^1$. To be precise, the toric data implies it is the space
\begin{equation}
  \P\bigl(\O(-2)\oplus\O\oplus\O\oplus\O\bigr),
\end{equation}
over $\P^1$. The short exact sequence
\begin{equation}
\xymatrix@1{
0\ar[r]&\O(-2)\ar[r]&\O(-1)^{\oplus2}\ar[r]&\O\ar[r]&0,
}
\end{equation}
on $\P^1$ shows that $\O(-1)^{\oplus2}$ may be viewed as a deformation
of $\O\oplus\O(-2)$. That is, we may construct a family of line
bundles over $\P^1$ where the central fibre is $\O\oplus\O(-2)$ and
all other fibres are $\O(-1)^{\oplus2}$. In the same way, there is a
three-dimensional space of first-order deformations of complex
structure that take
$\P\bigl(\O(-2)\oplus\O\oplus\O\oplus\O\bigr)$ into
$\P\bigl(\O(-1)^{\oplus2}\oplus\O\oplus\O\bigr)$.

So let $V_1$ be $\P\bigl(\O(-1)^{\oplus2}\oplus\O\oplus\O\bigr)$ which
we may define as a toric variety with homogeneous coordinate ring 
$R=\C[x_0,\ldots,x_5]$, 
\begin{equation}
 \Phi = \begin{pmatrix}
    1&1&1&1&0&0\\
    1&1&0&0&1&1
  \end{pmatrix},
\end{equation}
and $B=(x_0,x_1,x_4,x_5)(x_2,x_3)$. Let $X_1$ be a \CY\
hypersurface. For a specific complex structure one might consider
the defining equation
\begin{equation}
  W_1 = x_0^4+x_1^4 + x_2^4x_4^4+x_2^4x_5^4+x_3^4x_4^4+\lambda
  x_3^4x_5^4,  \label{eq:X1F}
\end{equation}
where $\lambda$ is a generic complex number (not equal to one or else
the threefold is singular). 

From what we have said, $X_0$ and $X_1$ are deformation
equivalent.\footnote{One can also show that the A-model
chiral rings of the $X_0$ and $X_1$ GLSMs are isomorphic:  these
models are describing the same moduli space of (2,2) SCFTs.}
 Putting $x_4=x_5=0$ in $X_1$ forces $x_0^4+x_1^4=0$ and thus
yields 4 rational curves. These 4 rational curves in $X_1$ are what
remains of the genus 3 curve times $\P^1$ in $X_0$ after switching on
any of the three non-polynomial deformations of $X_0$.

Note that $X_1$ exhibits all 86 deformations of complex structure as
polynomial deformations.

Let us compute $H^1(X_1,\End(T))$. The spectral sequence yields
\begin{equation}
\begin{matrix}\vspace{20mm}\\E_1^{p,q}:\end{matrix}
\begin{xy}
\xymatrix@C=5mm@R=5mm{
  \smc{208}\ar[r]&\smc{420}\ar[r]&\smc{33}\\
  &\scriptstyle0\ar[r]&\smc{8}\\
  &&\smc{8}\ar[r]&\scriptstyle0\\
  &&\smc{33}\ar[r]&\smc{420}\ar[r]^{d_1}&\smc{208}
} 
\save="x"!LD+<-3mm,0pt>;"x"!RD+<0pt,0pt>**\dir{-}?>*\dir{>}\restore
\save="x"!LD+<23mm,-3mm>;"x"!LU+<23mm,2mm>**\dir{-}?>*\dir{>}\restore
\save!RD+<0mm,-4mm>*{p}\restore
\save!CL+<21mm,0mm>*{q}\restore
\end{xy}
\end{equation}

The map $d_1$ fails to be surjective, similarly to the $X_0$ case. The
next stage of the spectral sequence is
\begin{equation}
\begin{matrix}\vspace{20mm}\\E_2^{p,q}:\end{matrix}
\begin{xy}
\xymatrix@C=5mm@R=5mm{
  \smc{}\ar[rrd]&\smc{181}&\smc{}\\
  &\scriptstyle0&\smc{8}\\
  &&\smc{8}\ar[rrd]&\scriptstyle0\\
  &&\smc{}&\smc{181}&\smc{}
} 
\save="x"!LD+<-3mm,0pt>;"x"!RD+<0pt,0pt>**\dir{-}?>*\dir{>}\restore
\save="x"!LD+<22mm,-3mm>;"x"!LU+<22mm,2mm>**\dir{-}?>*\dir{>}\restore
\save!RD+<0mm,-4mm>*{p}\restore
\save!CL+<20mm,0mm>*{q}\restore
\end{xy}
\end{equation}
The $d_2$ maps in this spectral sequence may or may not be zero
depending on the precise complex structure. The computation is very
similar to that for $X_0$ in the appendix.
The result is that $d_2$ is zero for the specific equation
(\ref{eq:X1F}) but becomes nonzero for a generic defining equation.
Thus
\begin{equation}
\dim H^1(X_1,\End(T))=\begin{cases}
  190\quad\hbox{for the hypersurface (\ref{eq:X1F})}\\
  188\quad\hbox{generically}
\end{cases}
\end{equation}

This fits in very nicely with the picture for $X_0$. Once again we
have a generic value of 188 for $\dim H^1(X,\End(T))$ but this can
increase for specific complex structures. $X_1$ is ``more generic''
than $X_0$ and cannot achieve as large a value for $\dim
H^1(X,\End(T))$.  The numbers of GLSM deformations for $X_0$
and $X_1$ are, respectively, $179$ and $180$.

\subsubsection{The Orbifold} \label{sss:oct-orb}

$X_{\textrm{orb}}$ is the singular octic in the unresolved weighted
projective space $\P^4_{\{2,2,2,1,1\}}$. We may compute the untwisted
sector easily enough in a way analogous to section \ref{ss:orb7}. The
result is that
\begin{equation}
\begin{split}
  h^{1,1}_0&=1\\
  h^{2,1}_0&=83\\
  \dim H^1(X,\End(T))_0&=182,
\end{split}
\end{equation}
where the subscript 0 denotes the untwisted sector.

$X_{\textrm{orb}}$ exhibits a genus 3 curve $C$
of singularities and $\Gamma=\Z_2$.   We now have two twisted sectors
with $k=1$, where $\nu_1=0$ and $\nu_a=\ff{1}{2}$ for $a=2,3$.  Note that the
$k=1$ sectors are self-conjugate, and (\ref{eq:orbcharge}) shows that 
the twisted vacua are $\Z_2$-invariant.

In the $(k,s)=(1,0)$ (R) sector  we have $\nut_1=0,\
\nut_a=-\ff{1}{2}$.  $E_{(1,0)}=0$, and the vacuum charges are $(q_{(1,0)},\bar
q_{(1,0)}) = (-\ff{1}{2},-\ff{1}{2})$.  In this sector we have zero modes of
$\gamma_1$ and $\psi_1$ so that the ground state is fourfold
degenerate.  Letting $|1,0\ra$ denote the state annihilated by
$\gamma_1$ and $\psi_1$, we have the four ground states with their
associated charges and spins along $C$, $(q,\bar q;q_C)$
\begin{equation}
|1,0\ra_{(-1/2,-1/2;1)}\quad \gammab_1|1,0\ra_{(1/2,-1/2;0)}\quad
\bar\psi_1|1,0\ra_{(-1/2,1/2;0)}\quad
\gammab_1\bar\psi_1|1,0\ra_{(1/2,1/2;-1)}\ .
\end{equation}
All of these are $\Gamma$-invariant and after GSO projection they lead
to massless chiral supermultiplets with
$\SO(10)\times\GU(1)_L\times\GU(1)_C$ transformation properties
$\brep{16}_{-1/2; 0}\oplus\rep{16}_{1/2;1}$ along with their
antichiral conjugates.

In the $(1,1)$ (NS) sector we have $\nut_1=-\ff{1}{2},\ \nut_a=0$.  Here
$E_{(1,1)} = -\ff{1}{2}$, and the charges are $(q_{(1,1)},\bar q_{(1,1)}) =
(-1,-\ff{1}{2})$.   Here there are zero modes of $\gamma_a$ as well as
$\bar\psi_1$ and hence the ground state is eightfold degenerate. With
our usual notation we have 
\begin{align}
&|1,1\ra_{(-1,-1/2;0)}&  &\gammab_2\gammab_3|1,1\ra_{(1,-1/2;1)}&
&\gammab_a|1,1\ra_{(0,-1/2;1/2)}\nonumber\\ 
&\bar\psi_1|1,1\ra_{(-1,1/2;-1)}&
&\bar\psi_1\gammab_2\gammab_3|1,1\ra_{(1,1/2;0)}&
&\bar\psi_1\gammab_a|1,1\ra_{(0,1/2;-1/2)}
\end{align}
The first line will lead to chiral multiplets, the second to
antichiral (after acting with left-moving excitations).  The two
states listed last on each line are $\Gamma$-odd, the others
$\Gamma$-even.  
The $\Gamma$-even ground states will lead to massless chiral
multiplets transforming as $\rep{10}_{-1;0}\oplus\rep{10}_{1;1}$ and
their antichiral conjugates.  Acting with the supercharges we find the
associated singlets
\begin{align}\label{eq:Gmoduli}
G^+|1,1\ra &= \sum_a x_a\gammab_a|1,1\ra_{(0,-1/2;0)}& 
&\rep{1}_{0;0}\nonumber\\
G^-\gammab_2\gammab_3|1,1\ra &=
\left(\rho_2\gammab_3-\rho_3\gammab_2\right)|1,1\ra_{(0,-1/2;1)}&
&\rep{1}_{0;1}\ .
\end{align}

Internal excitations lead to additional GSO and $\Gamma$ invariant
chiral states as listed below, with their
$\SO(10)\times\GU(1)_L\times\GU(1)_C$ transformation properties, as
well as the antichiral conjugates
\begin{align}
&\gamma_1|1,1\ra &\rep{1}_{-2;1}\nonumber\\
&\gammab_1\gammab_2\gammab_3|1,1\ra &\rep{1}_{2;0}\nonumber\\
&\gamma_1\gammab_2\gammab_3|1,1\ra &\rep{1}_{0;2}\nonumber\\
&\gammab_1|1,1\ra &\rep{1}_{0;-1}\\
&x_a\gammab_b |1,1\ra_4 &\rep{1}_{0;0}\nonumber\\
&\rho_a\gammab_b|1,1\ra_4 &\rep{1}_{0;1}\nonumber
\end{align}
Note that $x_a,\rho_a$ have spins $-\ff{1}{2},\ff{1}{2}$, respectively. 

Collecting all of this together into representations of $\GE_6\times
U(1)_C$ we find that the chiral multiplets fill out
\begin{equation}
\rep{27}_0\oplus\brep{27}_1\oplus\rep{1}_{-1}\oplus
\rep{1}_2\oplus
\left(\rep{1}_{1}\right)^{\oplus 4} \oplus \left(\rep{1}_0\right)^{\oplus 4}
\end{equation}
with the antichiral conjugates as expected.  Note that one of the
states in $\left(\rep{1}_0\right)^{\oplus 4}$ is related by (\ref{eq:Gmoduli})
 to the
$\rep{27}_0$ and one of the states  in
$\left(\rep{1}_1\right)^{\oplus 4}$ is related to the $\brep{27}_1$.

When we compactify on $C$ we now find that massless chiral multiplets
are given by holomorphic sections of $\Omega^{\otimes q_C}$.  Riemann-Roch and
vanishing theorems give
\begin{equation}
  h^0(\Omega^{\otimes q_C}) = \begin{cases}
    0&q_C<0\\
    1&q_C=0\\
    g&q_C=1\\
    (2q_C-1)(g-1)&q_C>1.
\end{cases}
\end{equation}

For $q_C=0$ we thus predict from
the twisted sector one massless $\rep{27}$ and the single associated
K\"ahler modulus, as well as $3$ additional singlets regardless of the
genus of $C$.

Putting $g=3$ for the case at hand we predict, from $q_C=1$ states, $3$
massless $\brep{27}$s with the associated $3$ complex structure moduli,
as well as an additional $9$ singlets.  We also find $6$ massless
singlets from the $q_C=2$ states. 

Adding together the untwisted and twisted states we obtain
\begin{equation}
\begin{split}
  h^{1,1}(X_{\textrm{orb}}) &= 1 + 1 = 2,\\
  h^{2,1}(X_{\textrm{orb}})  &= 83 + 3 = 86
\end{split}
\end{equation}
as expected. More interestingly, we have a grand total of 288 singlets.
That is, we predict 200 singlets associated with $H^1(\End(T))$.
{\em This total agrees with the
massless spectrum for large radius phase for the Fermat
complex structure.}

So we have agreement with the large radius (and \LG\ phase --- as we
see shortly) only for particular values of the complex structure of
$X_0$. In particular, our orbifold computation seems, at first sight,
not to depend at all on the value of the complex structure of
$X_0$. How can we resolve this discrepancy?

We do not know the full resolution of this question, but we can make
the following observations. Our analysis in terms of free fields of
the twisted sector effectively assumes that we were analyzing the
normal bundle of $C$ in $X_0$ rather than $X_0$ itself. The deviations
of the geometry away from the normal bundle as we move away from $C$
must introduce corrections that we have ignored so far. 

Which states counted above do we believe are reliably massless?  The
fields in nontrivial $\GE_6$ representations, along with the
associated K\"ahler and complex structure moduli, are of course
protected by the $(2,2)$ supersymmetry; in addition, the $6$ singlet
states coming from $q_C=2$ involve only excitations along $C$ and we
do not expect them to be sensitive to the details of the structure of
$X$ away from the curve.  The $12$ additional singlet states coming
from $q_C=0,1$ involve excitations in the directions transverse to the
curve. There is nothing to stop these states being lifted by the
additional interactions introduced upon varying the
superpotential. Thus we might reasonably expect there to be 12 fewer
singlets for a generic complex structure.
This agrees perfectly with the large radius picture.

\subsubsection{The \LG\ locus}
Using the GLSM, it is easy to describe the \LG\ point of the GLSM
for $X_0$.
Integrating out the $x_5$ and $p$ fields leads to a theory with fields
$x_0, \ldots, x_4$, with $\alpha_{0,1,2} = \ff{1}{4}$, and $\alpha_{3,4} =
\ff{1}{8}$.  Assigning weights $[2,2,2,1,1]$ to the fields, the \LG\
superpotential is a degree $8$ polynomial in $5$ variables which we
denote 
\begin{equation}
\Wt(x_0\ldots x_4)\equiv W|_{x_5=1}\ .
\end{equation}
We will distinguish the fields with different $\alpha_i$ by taking
indices $I,J = 0,1,2$, and $a,b = 3,4$.  

\begin{table}[!t]
\vspace{-10mm}
\[
\xymatrix@C=10mm@R=5mm{
  \bar q=-\ff32&\bar q=-\ff12&\bar q=\ff12\\
{\begin{matrix}
\gamma_a\gammab_b|1\ra_{4}\\\oplus\\
\gamma_I\gammab_J|1\ra_9\oplus
H^{aJ}_{[1]}\gamma_a\gammab_J|1\ra_{12}\\\oplus\\
G_{[2]}^I\rho_I|1\ra_{18}\oplus G_{[1]}^a \rho_a|1\ra_4\end{matrix}}
\ar[r]^-\Qb&
{\begin{matrix}
H_{[6]}^I \gamma_I |1\ra_{150}\\\oplus\\
H_{[7]}^a\gamma_a|1\ra_{140}\end{matrix}}\\
&{\begin{matrix}
K_{[2]}^I \gammab_I|3\ra_{18}\\\oplus\\
K_{[1]}^a\gammab_a|3\ra_4\end{matrix}}\\
&\rho_I|5\ra_3\\
&\rho_I \rho_J\rho_K \rho_L \gammab_3 \gammab_4 |7\ra_{15}\ar[r]^-{\Qb}&
{\begin{matrix}
\rho_I\gamma_J \gammab_3\gammab_4 |7\ra_9\oplus
x_a \rho_J \gammab_b |7\ra_{12}\\\oplus\\
\rho_a\gammab_b|7\ra_4\oplus
\gammab_I|7\ra_{3}\end{matrix}}\\
&{\begin{matrix}
x_I \gamma_3\gamma_4\gammab_J |9\ra_9\oplus
x_I \rho_a \gamma_b |9\ra_{12}\\\oplus\\
x_a\gamma_b|9\ra_4\oplus
\gamma_I|9\ra_{3}\end{matrix}}\ar[r]^-{\Qb}&
F_{[4]} \gamma_3 \gamma_4 |9\ra_{15}\\
&&x_I|11\ra_3\\
&&\vdots
}
\]
\caption{\LG\ states for the octic.}  \label{t:LG8}
\end{table}

With these numbers in hand, we classify the zero energy states with
$q=0$ as shown in table \ref{t:LG8}.

For generic $\Wt$ we will have $244$ singlets from $k=1$, while for
Fermat we find $248$.  As in the example of the quintic, the physical
reason for the appearance of these extra singlets is just the Higgs
mechanism.

The $k=3$ and $k=5$ zero energy states clearly satisfy $\Qb = 0$, so that all of these
correspond to massless singlets.

The $k=9$ sector states are the CPT conjugates of the states at $k=7$
and the $\Qb$ complex is the transpose of the complex for $k=7$.  We
can thus equivalently study either sector, and since the structure at
$k=9$ more transparently reflects the description in the previous
subsection we choose this option.  The antichiral multiplets from this
sector (conjugate to the chiral multiplets at $k=7$) are determined by
the cokernel of $\Qb$.  Using $\nu_I = \ff{1}{8}$ and $\nu_a=\ff{9}{16}$ and
hence $\nut_I = -\ff{3}{8}, \nut_a=-\ff{15}{16}$, the map in this sector is
determined by the following terms
\begin{align}
\Qb_1 &= \gammab_I^\dagger \left(\p_I\Wt\right)_{\nut_I} =
\gammab_I^\dagger \left.\left(\p_I \Wt (x)\right)\right|_{x_3=x_4=0}\nonumber\\
\Qb_2 &= \gamma_a \left(\p_a \Wt\right)_{1+\nut_a} =
\gamma_a\left(\p_{ai}\Wt (x)_{-1/2}\right) \rho_i^\dagger =
\gamma_a\rho_b^\dagger \p_{ab} \Wt (x)|_{x_3=x_4=0} \nonumber \\
\end{align}
where we have used the moding information, and in the second line also
the absence of $\rho_I$ excitations at $\qb = -\ff{1}{2}$.

The first of these contributes the following nontrivial $\Qb$ action
\begin{equation}
\Qb_1 x_I\gamma_3\gamma_4\gammab_J|9\ra =
x_I\left.\left(\p_J \Wt (x)\right)\right|_{x_3=x_4=0}\gamma_3\gamma_4|9\ra\ ,
\end{equation}
so that the cokernel agrees with (\ref{eq:cokerdi}).  When $\Wt$ is in
Fermat form $\Qb_2=0$ and there is a six-dimensional space of
antichiral singlets at $\qb=\ff{1}{2}$, leaving $19$ chiral singlets at
$\qb=-\ff{1}{2}$.  
When we deform $\Wt$ away from the Fermat form the cokernel can
decrease.  For instance, adding the term 
\begin{equation}
\delta \Wt = \psi x_0x_1x_2x_3x_4
\end{equation}
leaves $\Qb_1$ unchanged but adds
\begin{equation}
\delta\Qb_2 = \psi\left(\gamma_3\rho_4^\dagger +
  \gamma_4\rho_3^\dagger\right) x_0x_1x_2
\end{equation}
which now acts nontrivially:
\begin{equation}
\delta\Qb_2 x_I\rho_a\gamma_b|9\ra = \pm
\psi x_I \delta_{ab} x_0x_1x_2\gamma_3\gamma_4|9\ra.
\end{equation}
 The three-dimensional image
of this is clearly independent of the image of $\Qb_1$ so this
deformation removes three additional pairs of states, leaving $16$
chiral and $3$ antichiral singlets at $k=9$.  For generic $\Wt$, $\Qb$
has no cokernel and the singlet spectrum is simply $13$ chiral
states. 

To see that this is identical\footnote{Assuming the higher
  differential $d_5$ in the appendix always vanishes.}  to the
calculation in the appendix note that in general
\begin{equation}
\begin{split}
\Qb_2 x_I\rho_a\gamma_b|9\ra &= \epsilon_{ab}
x_I\left.\left(\p_{34}\Wt\right)\right|_{x_3=x_4=0}\gamma_3\gamma_4|9\ra\\ 
&= x_I\epsilon_{ab}\left(\p_5 W\right)_{x_5=0}\gamma_3\gamma_4|9\ra
\end{split}
\end{equation}
using the gauge invariance of $W$.

The most obvious physical explanation for this lifting of states is
via a simple mass term in the superpotential of the effective theory.
The (anti)chiral states in $k=7$ have their chiral(antichiral)
conjugates in the $k=9$ sector, allowing for a mass term that depends
on the untwisted complex structure moduli.  

To summarize, we find $282=2+86+188+6$ singlets for generic $W$, while
the Fermat point has an additional $4+12$ massless singlets. This
agrees with the Gepner model value of 298.

\subsubsection{Mirror Symmetry and K\"ahler Deformations}

The singlet spectrum at the Gepner point in this model includes, in
addition to the ``cascade states'' which as expected are not lifted by
K\"ahler deformations, and the expected four additional singlets
lifted by the Higgs mechanism under any deformation, a total of 28
additional chiral singlets in the sectors $k=5,7,9$.  The Gepner model
here is $A_3^{\oplus 3}\oplus A_7^{\oplus 2}$.
The model enjoys
a discrete permutation symmetry $S_3\times\Z_2$ and the states form
representations of this.  This symmetry commutes with the quantum
symmetry associated to the Gepner orbifold so that states within each
twisted sector transform into each other.   As always, orbits under
the discrete symmetry are lifted together under the K\"ahler
deformations away from the Gepner point, both of which are invariant.

Explicitly, the 28 states comprise eight orbits of the discrete symmetry
as listed below.  The number in brackets 
indicates the size of the associated orbit.  The $3$ chiral states in the
$k=5$ sector form the orbit of
\begin{equation*}
S_1 =  \lqs{\pz1}{-3}2\lqs {\pz1}{-3}2\lqs {\pz0}{-4}2\lqs {\pz2}{-2}0\lqs {\pz2}{-2}0\Bigl|
\lqs 121\lqs 121\lqs 011\lqs 231\lqs 231\qquad [3]
\end{equation*}
The $6$ chiral states in the $k=7$ sector comprise the orbits of 
\begin{align*}
S_2 &= \lqs{\pz2}{-4}2\lqs {\pz0}{-6}2\lqs {\pz0}{-6}2\lqs {\pz2}{-4}2\lqs {\pz2}{-4}2\Bigl|
\lqs 131\lqs 011\lqs 011\lqs 231\lqs 231 & [3]\\
S_3 &= \lqs{\pz1}{-5}2\lqs {\pz1}{-5}2\lqs {\pz0}{-6}2\lqs {\pz2}{-4}2\lqs {\pz2}{-4}2\Bigl|
\lqs 121\lqs 121\lqs 011\lqs 231\lqs 231 & [3]\\
\end{align*}
The $19$ chiral states in the $k=9$ sector comprise the orbits of 
\begin{align*}
S_4 &= \lqs{\pz1}{-7}0\lqs 000\lqs 000\lqs {\pz4}{-4}0\lqs {\pz2}{-6}2\Bigl|
\lqs 121\lqs 111\lqs 011\lqs 451\lqs 231 & [6]\\
S_5 &= \lqs{\pz1}{-7}2\lqs 000\lqs 000\lqs {\pz3}{-5}2\lqs {\pz3}{-5}2\Bigl|
\lqs 121\lqs 011\lqs 011\lqs 341\lqs 341 & [3]\\
S_6 &= \lqs000\lqs 000\lqs 000\lqs {\pz5}{-3}0\lqs {\pz3}{-5}0\Bigl|
\lqs 011\lqs 011\lqs 011\lqs 561\lqs 341 & [2]\\
S_7 &= \lqs000\lqs 000\lqs 000\lqs {\pz4}{-4}0\lqs {\pz4}{-4}2\Bigl|
\lqs 011\lqs 011\lqs 011\lqs 451\lqs 451 & [2]\\
S_8 &= \lqs{\pz1}{-7}0\lqs 000\lqs 000\lqs {\pz3}{-5}2\lqs {\pz3}{-5}0\Bigl|
\lqs 121\lqs 011\lqs 011\lqs 341\lqs 341 & [6]\\
\end{align*}

The mirror is given by a $\Z_4^3$ orbifold, with
\begin{equation}
m^{(1)} = (1,0,0,0,-2)\quad m^{(2)} = (0,1,0,0,-2)\quad m^{(3)} =
(0,0,1,0,-2)\ .
\end{equation}
Table \ref{t:twistoc} shows the twisted sectors in which the mirrors
of each of these states appear.  Note that the $S_3$ permutation
permutes $t^{(a)}$ while the $\Z_2$ acts as $(k,t_a)\mapsto (k - 4\sum
t_a,t_a+2\sum t_a)$.
\begin{table}[t]\label{t:twistoc}
\begin{center}
\begin{tabular}{|c|c|c|}
\hline
$(k;t_1,t_2.t_3)$	&Orbits	& Singlets at Gepner point	\\ \hline
$(1;0,0,0)$		& $7[2]$			& $7$\\ \hline
$(1;1,0,0)$		& $4$			& $3$\\ \hline
$(1;3,3,2)$		& $1$			& $1$\\ \hline
$(3;3,3,3)$		& $6$			& $2$\\ \hline
$(13;3,2,2)$		& $4$			& $3$\\ \hline
$(15;0,2,2)$		& $2$			& $1$\\ \hline
$(15;2,1,1)$		& $5;8[2]$		& $3$\\ \hline
$(15;3,3,2)$		& $3$			& $3$\\ \hline
\end{tabular}
\end{center}
\caption{Twisted Sectors $(k;t)$ in which non-cascade singlets arise}
\end{table}

The mirror model has two untwisted (2,2) deformations corresponding to
the two toric K\"ahler deformations of the octic.  We can write the
general superpotential for the mirror model as
\begin{equation}
\widehat W = x_0^4 + x_1^4 + x_2^4 + x_3^8 + x_4^8 - 8\psi x_0 x_1 x_2
x_3 x_4 - 4\chi x_3^4 x_4^4\ .
\end{equation}
In each of the sectors in table \ref{t:twistoc} sectors we have computed
the $\Qb$ cohomology and the only one in which this changes when we
deform $\widehat W$ is the sector $(15;2,1,1)$ (and its three
permutations).  In this sector we have 
\begin{equation}\label{eq:nus15}
\nu = (\ff38,\ff18,\ff18,\ff{15}{16},\ff{15}{16});\qquad
\nut =(-\ff18,-\ff38,-\ff38,-\ff9{16},-\ff9{16})
\end{equation}
and a basis for the zero energy states at $q=0$ is
\begin{equation}
\xymatrix@C=40mm@R=5mm{
  \bar q=-\ff12&\bar q=\ff12\\
{\begin{matrix}
\gammab_I x_I\rho_3^3\rho_4^3|v\ra_3\\
\gammab_3\rho_3^2\rho_4^3|v\ra\\
\gammab_4\rho_3^3\rho_4^2|v\ra
\end{matrix}}
\ar[r]^-\Qb&
{\begin{matrix}
x_1^4\rho_3^3\rho_4^3|v\ra\\
x_2^4\rho_3^3\rho_4^3|v\ra\\
x_0x_1x_2\rho_3^2\rho_4^2|v\ra
\end{matrix}}
}
\label{eq:lifted}
\end{equation}
where $|v\ra$ is the twisted vacuum.

The map $\Qb$ is here given by the terms
$\Qb = \gammab_i^\dagger \left(\p_i\widehat W\right)_{\nut_i}$ and
inserting (\ref{eq:nus15}) we compute for our explicit superpotential
\begin{align*}
\left(\p_0\widehat W\right)_{-1/8} &=  12x_0^2\rho_0^\dagger\\
\left(\p_1\widehat W\right)_{-3/8} &=  4x_1^3\\
\left(\p_2\widehat W\right)_{-3/8} &=  4x_2^3\\
\left(\p_3\widehat W\right)_{-9/16} &= 8! x_3\rho_3^{\dagger 6}
 - 8\psi\left(x_1x_2x_4\rho_0^\dagger + x_0x_2x_4\rho_1^\dagger + 
x_0x_1x_4\rho_2^\dagger + x_0x_1x_2\rho_4^\dagger\right)\\
&- 4(4!)^2\chi\rho_3^{\dagger 2}\rho_4^{\dagger 3}\left(x_3\rho_4^\dagger
  + x_4\rho_3^{\dagger}\right)\\
\left(\p_4\widehat W\right)_{-9/16} &= 8! x_4\rho_4^{\dagger 6}
 - 8\psi\left(x_1x_2x_3\rho_0^\dagger + x_0x_2x_3\rho_1^\dagger + 
x_0x_1x_3\rho_2^\dagger + x_0x_1x_2\rho_3^\dagger\right)\\
&- 4(4!)^2\chi\rho_3^{\dagger 3}\rho_4^{\dagger 2}\left(x_3\rho_4^\dagger
  + x_4\rho_3^{\dagger}\right)
\end{align*}
At the Gepner point ($\psi=\chi=0$) we have 
\begin{equation}
\Qb \gammab_1 x_1\rho_3^3\rho_4^3|v\ra = 
4x_1^4\rho_3^3\rho_4^3|v\ra\qquad
\Qb \gammab_2 x_2\rho_3^3\rho_4^3|v\ra = 
4x_2^4\rho_3^3\rho_4^3|v\ra
\end{equation}
as the only nontrivial $\Qb$ action.  This leaves $3$ chiral singlet
states in this sector at the Gepner point, as observed.
When we deform, we find that in addition
\begin{equation}
\Qb \gammab_3\rho_3^2\rho_4^3|v\ra = 
\Qb \gammab_4\rho_3^3\rho_4^2|v\ra = 
-8\psi x_0x_1x_2\rho_3^2\rho_4^2|v\ra\ .
\end{equation}
Thus, in this sector (and in each of the two related to it by
permutations) as well as its conjugate sector (and its permutations)
one chiral state is lifted by the $\psi$ deformation.  
This agrees with the counting above.

\begin{table}[!t]
\renewcommand{\arraystretch}{1.3}
\begin{center}
\begin{tabular}{|c|r|}
\hline
K\"ahler form, untwisted&1\\
\hline
K\"ahler form, twisted&1\\
\hline
Complex Structure, untwisted&83\\
\hline
Complex Structure, twisted&3\\
\hline
$H^1(\End(T))$ generic untwisted&182\\
\hline
$H^1(\End(T))$ generic twisted&6\\
\hline
Extra singlets at special complex structure&12\\
\hline
Extra singlets at special K\"ahler form&6\\
\hline
Extra singlets associated to Gepner $\GU(1)^4$&4\\
\hhline{|=|=|}
Total&298\\
\hline
\end{tabular}
\end{center} \caption{Classification of the singlets in the Gepner
  model.} \label{t:Goct}
\end{table}

Let us conclude the analysis of the octic with a summary of the
counting of the singlets. The Gepner model has 298 singlets which may
be cataloged as in table \ref{t:Goct}. In this table we label states
``twisted'' or ``untwisted'' according to the orbifold picture of
section \ref{sss:oct-orb}.

\section*{Acknowledgments}

We wish to thank G.~Smith and S.~Mapes for access to a preview version
of \cite{Mac2:toric}. IVM is supported in part by the German-Israeli
Project cooperation (DIP H.52) and the German-Israeli Fund (GIF).  He
would also like to thank B.~Wurm for enlightening discussions and the
CGTP for hospitality while some of this
work was being completed.  We also thank BIRS and the organizers of
the workshop on (0,2) mirror symmetry for providing a great
opportunity to begin our elephant exploration.  This work was
partially supported by NSF grants DMS--0606578 and DMS--0905923.  Any
opinions, findings, and conclusions or recommendations expressed in
this material are those of the authors and do not necessarily reflect
the views of the National Science Foundation.

\appendix

\section{Spectral Sequence Computations} \label{s:sscomp}

\subsection{Definitions}

Let us first review how to compute the cohomology of a line
bundle $\O(\mathbf{q})$ on a toric variety $V$. We need to know this
in some detail in order to be able to compute the necessary
differentials in the spectral sequence in section \ref{ss:EndT}. The
method, using local cohomology, is generalized from Grothendieck
\cite{Grot:localcoh} in \cite{EMS:ToricCoh}. As we describe it, this
method is not completely optimized for efficiency, but for our
purposes it is not only the actual computation of $H^k(V,\O(\mathbf{q}))$
which is important, but also the explicit form of cochains which
represents it.

We wish to compute sheaf cohomology in terms of \v Cech
cohomology. The reasoning is exactly
analogous to the case of projective space as discussed in 
\cite{Hartshorne:} chapter III.5. Let the irrelevant ideal be $B=(m_1,m_2,\ldots,m_l)$,
where $m_i$ are monomials. Consider the open sets
\begin{equation}
  U_i = V-Z(m_i).
\end{equation}
These open sets form an open cover of $V$. 

The coordinate ring of $U_i$ is given by the localization $R_{m_i}$
consisting of functions $f/m_i^n$ for $f\in R$. Similarly $R_{m_i,m_j}
=(R_{m_i})_{m_j}$ is the coordinate ring on $U_{ij}=U_i\cap U_j$.
We then form the local cochain complex from
\begin{equation}
  C^i_B(R) = \bigoplus_{j_1<\ldots<j_i} R_{m_{j_1},\ldots,m_{j_i}}.
\end{equation}
The differential $d:C^i_B(R)\to C^{i+1}_B(R)$ is given by the obvious
inclusion map with some $(-1)^j$ factors to ensure $d^2=0$. We will
elucidate these signs below.

$R$ admits a {\em fine grading\/} valued in $\Z^n$, where the grade of
each homogeneous coordinate is a basis vector. Let a subscript $p$
denote this fine grading. Then \cite{EMS:ToricCoh}
\begin{equation}
  C^*_B(R)_p = \bigoplus_{\{J|\neg(p)\subset\supp(m_J)\}}\hspace{-30pt}\C,
\end{equation}
where $J$ is a subset of $\{1,\ldots,l\}$, $m_J$ is the least common
multiple of $m_j, j\in J$ and $\neg(p)$ is the subset of $\{1,\ldots,l\}$
corresponding to negative grades. The differential $d$ maps the $J$
component of $C^*_B(R)_p$ to the $J'$ component as zero unless
$J'=J\cup j$, in which case it is $(-1)^e$ where $j$ is in the $e$th
position of $J'$. The local cohomology groups $H^i_B(R)_p$ are defined
by the local cochain complex $C^*_B(R)_p$. Note, in particular, that
they only depend on $p$ via $\neg(p)$.

The definition of \v Cech cohomology then gives
\begin{equation}
  \check C^i(\{U_i\},\O)_p = C^{i+1}_B(R)_p, \quad\hbox{for $i\geq 0$}.
\end{equation}
We also have $C^0_B(R)_p = R_p$. It is a fact that the covering given
by $U_i$ is affine and thus sufficiently fine to give \v Cech
cohomology. It follows that
\begin{equation}
  H^i(V,\O)_p = H^{i+1}_B(R)_p, \quad\hbox{for $i\geq 1$},
\end{equation}
and that 
\begin{equation}
\begin{split}
  H^0(V,\O)_p &= R_p\\
  H^0_B(R) &= H^1_B(R) = 0.
\end{split}
\end{equation}

We now have an explicit method of computing the spectral sequence in
theorem \ref{th:m}. Following the notation of \cite{BT:}, it is based
on a double complex $K^{p,q}$, where the vector spaces $K^{p,q}$ have
localized Laurent monomials as a basis. The vertical maps in the
double complex are the \v Cech boundaries as explained above, and the
horizontal maps come from the complex representing the sheaf in question.

\subsection{The example}

Let $V$ be the resolved weighted projective space
$\P^4_{\{2,2,2,1,1\}}$. Here $R=\C[x_0,\ldots,x_5]$ and
$B=(x_0,x_1,x_2,x_5)(x_3,x_4)=(x_0x_3,x_1x_3,x_2x_3,x_5x_3,
x_0x_4,x_1x_4,x_2x_4,x_5x_4)$. Let $t$ denote the fine grading $(0,0,0,-1,-1,0)$.
Then $C^*_B(R)_t$ is given by (starting at position zero)
\begin{equation}
\xymatrix@1{
0\ar[r]&0\ar[r]&\C^{16}\ar[r]&\C^{48}\ar[r]&\C^{68}\ar[r]&
\C^{56}\ar[r]&\C^{28}\ar[r]&\C^8\ar[r]&\C^1.}
\end{equation}
It is exact in every place except $H^2_B(R)_t=\C$.
A generator of $H^2_B(R)_t$ is given by the same Laurent monomial
$x_3^{-1}x_4^{-1}$ in the 16 localizations $R_{m_i,m_j}$, where
$i=0,\ldots3$ and $j=4\ldots7$. One can also show\footnote{For a quick
  method see, for example, section 3 of \cite{Must:LocalCoh}.} that
$H^2_B(R)_p$ vanishes unless $\neg(p)=\neg(t)$.

We now have enough information to compute $H^1(V,\O(\mathbf{q})$ for
any $\mathbf{q}\in D$. The map from the fine grading to the
coarse grading, $D$, is given by the grades (\ref{eq:wts21}),
which we repeat here for convenience:
\begin{equation}
  \Phi = \begin{pmatrix} 0&0&0&1&1&-2\\
         1&1&1&0&0&1\end{pmatrix}. 
\end{equation}

The monomial $x_3^{-1}x_4^{-1}$ has charge $(-2,0)$, and it
follows that $H^1(V,\O(-2,0))=\C$. Similarly 
\begin{itemize}
\item $\dim H^1(V,\O(-2,1))=3$
with basis
$\{x_0x_3^{-1}x_4^{-1},x_1x_3^{-1}x_4^{-1},x_2x_3^{-1}x_4^{-1}\}$ and
\item $\dim H^1(V,\O(-3,1))=6$ with basis
\[\{x_0x_3^{-2}x_4^{-1},x_0x_3^{-1}x_4^{-2},
x_1x_3^{-2}x_4^{-1},x_1x_3^{-1}x_4^{-2},x_2x_3^{-2}x_4^{-1},x_2x_3^{-1}x_4^{-2}\}.\]
\end{itemize}

The first row of the spectral sequence has, for $d_1^{\,0,1}:E^{0,1}\to
E^{1,1}$:
\begin{equation}
\xymatrix@1{
{\begin{matrix}H^1(\O_X)^{\oplus 4}\\\oplus\\
\bigoplus_{i,j} H^1(\O_X(\mathbf{q}_i-\mathbf{q}_j))\\\oplus\\
H^1(\O_X)\end{matrix}}
\ar[r]^-{d_1^{\,0,1}}&
{\begin{matrix}\bigoplus_iH^1(\O_X(\mathbf{q}_i))^{\oplus2}\\\oplus\\
\bigoplus_iH^1(\O_X(\mathbf{Q}-\mathbf{q}_i))\end{matrix}}
}
\end{equation}
In our case this becomes
\begin{equation}
\xymatrix@1{
{\begin{matrix}H^1(\O(-2,0))^{\oplus3}\\\oplus\\
H^1(\O(-3,1))^{\oplus2}\end{matrix}}
\ar[r]^-{d_1^{\,0,1}}&
H^1(\O(-2,1))^{\oplus2}.
}
\end{equation}
Note that in each case here the restriction of $H^1(\O_V(\mathbf{q}))$
to $H^1(\O_X(\mathbf{q}))$ is the identity, and so we have dropped the
subscript from $\O$.

The map $d_1^{\,0,1}$ is induced from the map $E$ in (\ref{eq:TX}), which is
given by
\begin{equation}
\xymatrix@1@C=30mm{
\O^{\oplus2}\ar[r]^-{\left(\begin{smallmatrix}
0&x_0\\0&x_1\\0&x_2\\x_3&0\\x_4&0\\-2x_5&x_5\end{smallmatrix}\right)}&
{\begin{matrix}\O(0,1)^{\oplus3}\\\oplus\\
\O(1,0)^{\oplus2}\\\oplus\\\O(-2,1)\end{matrix}}.
}
\end{equation}
Using the Laurent monomial representatives of these cocycles we easily
obtain the map $d_1^{\,0,1}$:
\begin{equation}
  d_1^{\,0,1} = \left(\begin{smallmatrix}
0&0&0&1&0&0&0&0&0&0&0&0&1&0&0\\
0&0&0&0&1&0&0&0&0&0&0&0&0&1&0\\
0&0&0&0&0&1&0&0&0&0&0&0&0&0&1\\
1&0&0&0&0&0&0&0&0&0&0&0&0&0&0\\
0&1&0&0&0&0&0&0&0&0&0&0&0&0&0\\
0&0&1&0&0&0&0&0&0&0&0&0&0&0&0\end{smallmatrix}\right).\label{eq:d1matrix}
\end{equation}
The spectral sequence at the $E_1$ stage is given by
(\ref{eq:ssE121}). Now we have computed all the $d_1$ maps, and we obtain
\begin{equation}
\begin{matrix}\vspace{20mm}\\E_2^{p,q}:\end{matrix}
\begin{xy}
\xymatrix@C=5mm@R=5mm{
  \smc{6}\ar[rrd]_{d_2}&\smc{185}&\smc{}\\
  &&\smc{9}\\
  &&\smc{9}\ar[rrd]^{d_2}\\
  &&\smc{}&\smc{185}&\smc{6}
} 
\save="x"!LD+<-3mm,0pt>;"x"!RD+<0pt,0pt>**\dir{-}?>*\dir{>}\restore
\save="x"!LD+<23mm,-3mm>;"x"!LU+<23mm,2mm>**\dir{-}?>*\dir{>}\restore
\save!RD+<0mm,-4mm>*{p}\restore
\save!CL+<21mm,0mm>*{q}\restore
\end{xy}
\end{equation}

So now we need to compute the $d_2$ maps. Fortunately they are related
by Serre duality. Let us focus on $d_2^{\,0,1}:\C^9\to\C^6$.

The map $d_2$ will depend upon the defining equation $W$. Only certain
monomials terms will contribute to $d_2$. For purposes of argument we
concentrate first on a single monomial $x_0x_2^2x_3^2x_5$.

The process of computing $d_2$ can be quite formidable in a general
spectral sequence, but the explicit representatives of elements of
the double complex $K^{p,q}$ as Laurent monomials makes the procedure 
straight-forward. 
Consider, as an example, the monomial $x_0x_3^{-2}x_4^{-1}$ representing an element
of $H^1(\O(-3,1))$ in $K^{0,1}$. 

The map to $K^{1,1}$ has two contributions. First we map 
$\check C^1(\O(-3,1))$ to $\check C^1(\O(-2,1)$ by multiplying by $x_i\mathbf{q}_i$,
which in the case of interest amounts to multiplying by $x_4$. We also
map $\check C^1(\O(-3,1))$ to $\check C^1(\O(-1,4))$ by multiplying by
$\partial_5W$. Thus we map $x_0x_3^{-2}x_4^{-1}$ to
\begin{equation}
  \left(\frac{x_0}{x_3^2},\frac{x_0^2x_2^2}{x_4}\right)
    \in \check C^1(\O(-2,1))\oplus \check C^1(\O(-1,4)).
\end{equation}

This is a \v Cech coboundary, and we can chase it downwards as
follows.  Recall that these monomial representatives of
$\check C^1(\O(\mathbf{q}))$ are really 16 copies of the same monomials under
the localization $R_{m_i,m_j}$. Computing the chain complex
$C^*_B(R)_{(0,0,0,-1,0,0)}$, we see that $x_0x_3^{-2}$ lies in the
coboundary of the 4 copies of the monomial localized to $R_{m_i}$,
where $i=0,\ldots,3$. Similarly $x_0^2x_2^2x_4^{-1}$ lies in the
coboundary of the 4 copies of the negated monomial localized to $R_{m_j}$,
where $i=4,\ldots,7$.

Finally we apply $d_1$ to push our element to $K^{0,2}$.
$x_0x_3^{-2}$ is multiplied by $\partial_5W$, while
$x_0^2x_2^2x_4^{-1}$ is multiplied by $-x_4$. Paying attention to
signs, the result is $x_0^2x_2^2$ in both cases. That is, we have a \v
Cech cochain which takes the value $x_0^2x_2^2$ in {\em all eight\/}
patches $R_{m_k}$, $k=1,\ldots,8$.

The fact this has the same value in all 8 patches means that this is
the localization of a monomial in $R$ itself. This had to be, since
$H^1_B(R)=0$, and allows us to interpret $x_0^2x_2^2$ as an element of
$H^0(\O(\mathbf{Q}))$.  The computation of $d_2$ can be
summarized in the following diagram:
\begin{equation}
\xymatrix@C=25mm{
\left(\frac{x_0}{x_3^2x_4}\right)_{ij}\ar[r]^-{\left(\begin{smallmatrix}
x_4\\x_0x_2^2x_3^2\end{smallmatrix}\right)}&
\left(\frac{x_0}{x_3^2}\right)_{ij},\left(\frac{x_0^2x_2^2}{x_4}\right)_{ij}\\
&\left(\frac{x_0}{x_3^2}\right)_i,-\left(\frac{x_0^2x_2^2}{x_4}\right)_j
\ar[u]^d\ar[r]^-{(x_0x_2^2x_3^2,\,-x_4)}&\left(x_0^2x_2^2\right)_k\\
&&x_0^2x_2^2\ar[u]^d
}
\end{equation}
Thus, $d_2$ is nonzero.

Given a generic defining equation $W$ with all possible
monomials, this $d_2$ map is surjective and the sequence degenerates at
the $E_3$ term:
\begin{equation}
\begin{matrix}\vspace{20mm}\\E_3^{p,q}:\end{matrix}
\begin{xy}
\xymatrix@C=5mm@R=5mm{
  &\smc{185}&\smc{}\\
  &&\smc{3}\\
  &&\smc{3}\\\
  &&\smc{}&\smc{185}&{\phantom{\smc{6}}}
} 
\save="x"!LD+<-3mm,0pt>;"x"!RD+<0pt,0pt>**\dir{-}?>*\dir{>}\restore
\save="x"!LD+<21mm,-3mm>;"x"!LU+<21mm,2mm>**\dir{-}?>*\dir{>}\restore
\save!RD+<0mm,-4mm>*{p}\restore
\save!CL+<19mm,0mm>*{q}\restore
\end{xy}
\end{equation}
So $\dim H^1(X,\End(T))=188$.

However, suppose we pick the Fermat form of the defining equation
$W$. Studying the above computation of $d_2$, it is clear that we can
never hit monomials in $K^{0,2}$ of the form $x_0^2x_2^2$, etc. Thus,
the $d_2$ maps are actually zero. Now we go to the $E_4$ stage:
\begin{equation}
\begin{matrix}\vspace{20mm}\\E_4^{p,q}:\end{matrix}
\begin{xy}
\xymatrix@C=5mm@R=5mm{
  \smc{6}\ar[rrrrddd]^(0.7){d_4}&\smc{185}&\smc{}\\
  &&\smc{9}\\
  &&\smc{9}\\
  &&\smc{}&\smc{185}&\smc{6}
} 
\save="x"!LD+<-3mm,0pt>;"x"!RD+<0pt,0pt>**\dir{-}?>*\dir{>}\restore
\save="x"!LD+<23mm,-3mm>;"x"!LU+<23mm,2mm>**\dir{-}?>*\dir{>}\restore
\save!RD+<0mm,-4mm>*{p}\restore
\save!CL+<21mm,0mm>*{q}\restore
\end{xy}
\end{equation}

The source of the $d_4$ map is a $\C^6$ subspace of
$H^3(\O_X(-\mathbf{Q}))^{\oplus 2}$. This third cohomology group is {\em not\/}
the isomorphic image of $H^3(\O_V(-\mathbf{Q}))$ under restriction, and
so we need to work a little harder to describe everything in terms of
cohomology of the toric variety and thus local cohomology. To this end
we may use the short exact sequence
\begin{equation}
\xymatrix@1{
0\ar[r]&\O_V(-\mathbf{Q})\ar[r]&\O_V\ar[r]&\O_X\ar[r]&0,
}
\end{equation}
to write $\O_X$ in terms of $\O_V$. By using mapping cones, we may write
the complex (\ref{eq:TX}) representing the tangent sheaf as
\begin{equation}
\xymatrix@1@C=23mm{
\O_V(-\mathbf{Q})^{\oplus r}\ar[r]^-{\left(\begin{smallmatrix}
W\id_r\\x_i\mathbf{q}_i\end{smallmatrix}\right)}&
{\begin{matrix}\O_V^{\oplus r}\\\oplus\\
\bigoplus_i\O_V(\mathbf{q}_i-\mathbf{Q})\end{matrix}}
\ar[r]^-{\left(\begin{smallmatrix}
x_i\mathbf{q}_i&-W\id_n\\\mathbf{Q}&\partial_iW\end{smallmatrix}\right)}&
\poso{{\begin{matrix}\bigoplus_i\O_V(\mathbf{q}_i)\\\oplus\\\O_V\end{matrix}}}
\ar[r]^-{\left(\begin{smallmatrix}\partial_iW&W\end{smallmatrix}\right)}&
\O_V(\mathbf{Q})
},
\end{equation}
and we can take the $\sHom$ of this complex into itself to form a
complex for $\End(T)$. The $H^3(\O_X(-\mathbf{Q}))^{\oplus 2}$ from above now
manifests itself as $H^4(\O_V(-2\mathbf{Q}))^{\oplus 2}$, and the
interesting map appears in $E_5$ as\footnote{This map is really only
  defined on a subspace of the term on the left and a quotient of the
  space on the right, since we are taking cohomology at the earlier
  stages of the spectral sequence.}
\begin{equation}
  d_5:H^4(\O_V(-2\mathbf{Q}))^{\oplus 2}\to H^0(\O_V(\mathbf{Q}))^{\oplus 2}.
\end{equation}
With a little organization one can show, for example, that the Fermat
form of $W$ cannot give rise to a nonzero map.\footnote{This $d_5$ map
  is a quite tedious to calculate but appears to be zero for any $W$.}  Thus,
in this case, the spectral sequence degenerates at the $E_2$ stage again, and
now $\dim H^1(X,\End(T))=185+9+6=200$. That is we have 12 extra states
compared to the generic hypersurface.


\end{document}